\documentclass[twocolumn]{aastex6}
\usepackage{natbib}
\begin{document}
\title{Formation and eruption of a flux rope from the sigmoid active region NOAA 11719 and associated M6.5 flare: A multi-wavelength study}

\author{Bhuwan Joshi}
\author{Upendra Kushwaha}
\affil{Udaipur Solar Observatory, Physical Research Laboratory, Udaipur 313001, India}


\author{Astrid M. Veronig}
\affil{Kanzelh\"ohe Observatory/Institute of Physics, University of Graz, Universit$\ddot{a}$tsplatz 5, A-8010 Graz, Austria}

\author{Sajal Kumar Dhara}
\affil{Udaipur Solar Observatory, Physical Research Laboratory, Udaipur 313001, India}

\author{A. Shanmugaraju}
\affil{Department of Physics, Arul Anandhar College, Karumathur, Tamilnadu 625514, India}

\author{Yong-Jae Moon}
\affil{School of Space Research, Kyung Hee University, Yongin, Gyeonggi-Do, 446-701, Korea}

\begin{abstract}
We investigate the formation, activation and eruption of a flux rope from the sigmoid active region NOAA 11719 by analyzing E(UV), X-ray and radio measurements. During the pre-eruption period of $\sim$7~hours, the AIA 94~\AA~images reveal the emergence of a coronal sigmoid through the interaction between two J-shaped bundles of loops which proceeds with multiple episodes of coronal loop brightenings and significant variations in the magnetic flux through the photosphere. These observations imply that repetitive magnetic reconnections likely play a key role in the formation of the sigmoidal flux rope in the corona and also contribute toward sustaining the temperature of the flux rope higher than the ambient coronal structures. Notably, the formation of the sigmoid is associated with the fast morphological evolution of an S-shaped filament channel in the chromosphere. The sigmoid activates toward eruption with the ascend of a large flux rope in the corona which is preceded by the decrease of photospheric magnetic flux through the core flaring region suggesting {\it tether-cutting} reconnection as a possible triggering mechanism. The flux rope eruption results in a two-ribbon M6.5 flare with a prolonged rise phase of $\sim$21 min. The flare exhibits significant deviation from the {\it standard flare model} in the early rise phase during which a pair of J-shaped flare ribbons form and apparently exhibit converging motions parallel to the polarity inversion line which is further confirmed by the motions of HXR footpoint sources. In the later stages, the flare follows the {\it standard flare model} and the source region undergoes a complete {\it sigmoid-to-arcade} transformation.
\end{abstract}


\section{Introduction}
Coronal mass ejections (CMEs) immensely affect space weather phenomena. Thus a major objective of research in solar physics in recent times has been to explore the source region characteristics of CMEs. Through these efforts, some conditions have been recognized that are favorable for eruptions. In this regard, the appearance of a ``sigmoid" in the active region corona is considered as an important precursor of CMEs. It is widely accepted that sheared and twisted coronal fields associated with sigmoids can store a large amount of free magnetic energy which is ultimately released during the CME. 

Sigmoids were first identified by the Yohkoh Soft X-ray Telescope (SXT) as large regions producing enhanced soft X-ray emission having S-shaped (or inverse S-shaped) morphology \citep{Rust1996, Manoharan1996,Pevtsov1996,Sterling1997,Moore2001}. Using SXT data, \cite{Hudson1998} studied the source region of several halo CMEs and found that, for the majority of events, the source region exhibited a characteristics pattern in which pre-eruption sigmoids turned into loop arcades following the passage of a CME \citep[see also][]{Sterling2000}. Using a large data set of SXT images, \cite{Canfield1999} classified solar active regions in sigmoidal and non-sigmoidal categories and found that the former type of activity centers are more likely to be eruptive than the other ones. According to morphology and evolution time-scale, sigmoids can be classified into two groups: transient and persistent \citep{Gibson2006}. Transient sigmoids brighten up only for a short period of time, usually just before the eruption \citep[see, e.g., ][]{Pevtsov1996}. They tend to be more well defined in the form of apparently a single, sigmoid loop. Notably, observations reveal that many sigmoids have the shape of two J's or elbows, which together form the S-shape of the sigmoid \citep[see, e.g.,][]{Moore2001}. Persistent sigmoids present much intricate morphology in which many discrete sheared loops collectively form a sigmoidal structure. They are long-lived features that sustain for considerably longer time than the transient sigmoids (from days to weeks)\citep[see, e.g., ][]{McKenzie2008}.  

Underneath the twisted coronal soft X-ray (SXR) structures, filament channels are frequently observed in H$\alpha$ observations \citep{Pevtsov1996, Pevtsov2002, Gibson2002}. Although the sigmoid-to-arcade evolution is quite dramatic, the underlying filament may or may not show significant changes with the sigmoid eruption \citep{Pevtsov2002}. Further, sigmoids can also develop over decayed active regions showing weak and dispersed distribution of magnetic flux \citep{Glover2001}. We believe that there is some coupling between these two structures (sigmoid and filament) although observations at these two channels (SXR and H$\alpha$) correspond to the hottest and the coldest material associated with the sigmoidal regions. Therefore, it is essential to probe what happens in-between these two layers and temperature regions during the formation and disruption stages of the sigmoids. This objective can be accomplished by analyzing suitable multi-wavelength data sets. Recent studies indicate that sigmoidal structures are visible in a wider range of temperature \citep{Liu2007}. Finally, we need to understand how the overlying coronal and chromospheric structures are related to the underlying magnetic field evolution through the photosphere.

The emergence or formation of magnetic flux ropes in solar active regions and their subsequent eruption has been recognized as the key component of the sigmoid-to-arcade evolution process \citep[see, e.g.,][]{Titov1999,Kliem2004, Archontis2009,Chatterjee2013,Schmieder2015,Jiang2016,Kumar2016}. Comtemporary observations, taken from multiple EUV channels of AIA, indeed provide evidence toward the existence and activation of the hot flux rope in the active region corona \citep[see, e.g.,][]{Cheng2011,ChengX2013_two_MFR, Kumar2014,ChenB2014, ChengX2016}. Further, the comparison between the kinematic evolution of the flux rope and associated CME reveals that the hot flux rope acts as the earliest signature of the CME \citep{Cheng2013_MRF_CME,Cheng2014_MFR_evol}.

\begin{figure*}
\epsscale{0.7}
\plotone{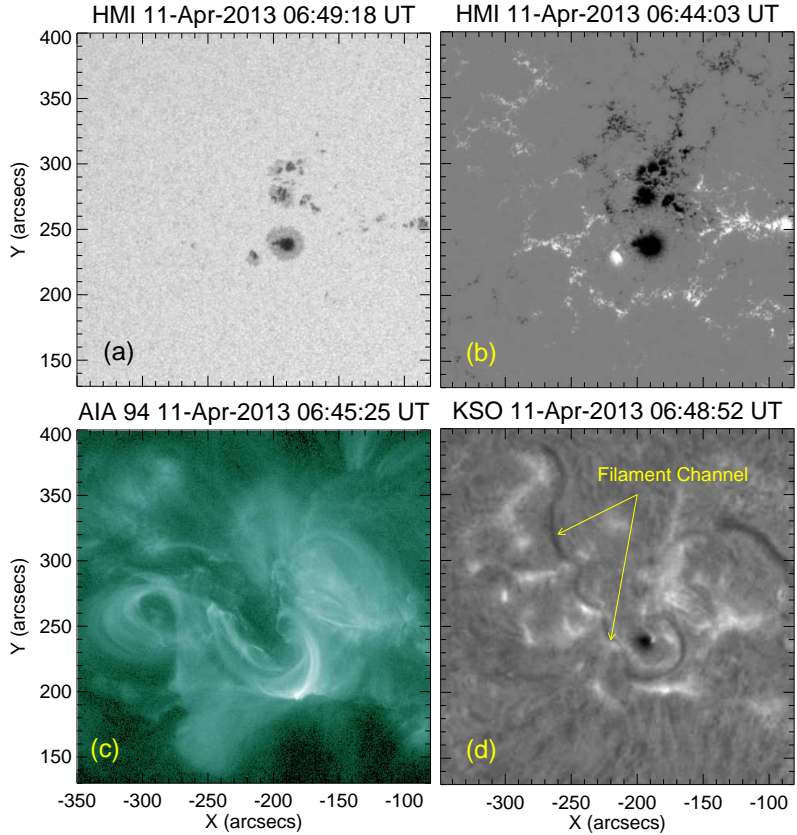}
\caption{Multi-wavelength view of active region 11719 on 2013 April 11. Panel (a): White light picture of the active region showing the distribution of sunspots. Panel (b): HMI line-of-sight magnetogram presenting the magnetic flux distribution of the active region. Panel (c): AIA 94 \AA~image of the pre-flare phase showing a sigmoidal structure.  Panel (d): KSO H$\alpha$ image showing a long filament channel (marked with arrows) along the polarity inversion line.}
\label{fig1}
\end{figure*}

In this study, we present a comprehensive multi-wavelength analysis of the morphological evolution and eruption of the sigmoidal active region NOAA 11719 on 2013 April 11. We discuss the dramatic evolution of this active region over a period of nine hours (00:00--09:00~UT). In this period, an extreme ultraviolet (EUV) sigmoid structure emerged through the interactions between two J-shaped bundles of loops that involves multiple events of localized energy release and photospheric flux changes. Subsequently, we observe the ascend and eruption of a flux rope from this sigmoidal region. This study is preliminary based on multi-channel E(UV) imaging taken from Atmospheric Imaging Assembly \citep[AIA;][]{Lemen2012} on board the Solar Dynamics Observatory (SDO) having unprecedented spatial and temporal resolutions. Notably, the evolution of the sigmoid was observed at AIA 94~\AA~ images which implies that the structure was comprised of very high temperature plasma ($\sim$6~MK). During the eruption, we observed a large M6.5 flare (SOL2013-04-11) which is characterized by a prolonged SXR rise phase of $\sim$21 min. It is striking that the flare evolution during the early rise phase significantly deviates from the standard flare model. The H$\alpha$ observations from Kanzelh\"{o}he Observatory \citep[KSO;][]{Potzi2015} revealed that a long active region filament existed below the coronal sigmoid which partially erupted during the M6.5 flare and caused a large two-ribbon flare in the chromosphere. The temporal and spatial evolution of hard X-ray (HXR) emission during the M6.5 flare was studied using multi-band X-ray time-profiles and images obtained from the Reuven Ramaty High Energy Solar Spectroscopic Imager \citep[RHESSI;][]{LinRP2002}. Detailed comparison of chromospheric and coronal activities during the sigmoid evolution with the changes in photospheric magnetic flux was undertaken to investigate the triggering mechanism involved in  this eruption. For magnetic field measurements, we have analyzed longitudinal magnetograms from the Helioseismic Magnetic Imager \citep[HMI;][]{Schou2012} on board SDO. These multi-wavelength observations are further supplemented by radio dynamic spectra obtained from the HiRAS radio spectrograph. We present an observational overview of the activities in Section~\ref{sec_obs}. The analysis and observational results are presented in Section~\ref{sec_analysis}. We interpret our results and emphasize the uniqueness of this work in Section~\ref{sec_discuss}. Summary of this study is given in Section~\ref{sec_conclusion}. 

\section{Overview of observations}
\label{sec_obs}

\begin{figure*}
\epsscale{0.7}
\plotone{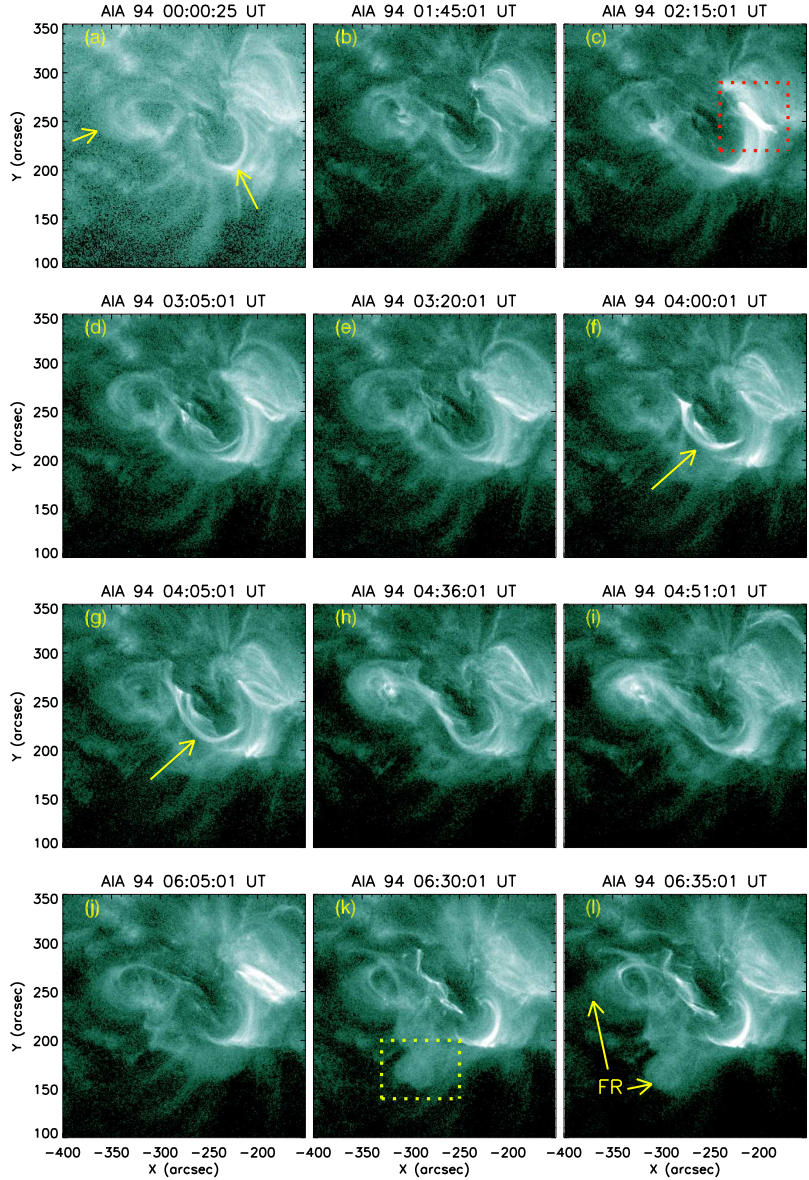}
\caption{Series of AIA 94 \AA~images showing the development of the sigmoidal structure in active region NOAA 11719 during the pre-eruption phase. Note the ascend of the flux rope (FR) from the active region (marked with a dotted-box region and arrows in panels (k) and (l), respectively).}
\label{Sigmoid_evolution}
\end{figure*}

In Figure \ref{fig1}, we present a multi-wavelength view of the active region NOAA 11719 at white light (WL), 94~\AA, and H$ \alpha $ wavelengths to show distribution of sunspots and associated coronal features. The WL image clearly indicates that the active region consists of several small-to-intermediate sized sunspots with the largest one possessing negative polarity (cf. Figures \ref{fig1}(a) and (b)). It is interesting to note that most of the prominent sunspots are of negative polarity. Further, there is scarcity of sunspots exhibiting positive polarity while the positive flux region is dispersed over a larger area (Figure \ref{fig1}(b)). The overall photospheric flux distribution suggests a $ \beta \gamma $ magnetic configuration of the AR. An inverse S-shaped structure (i.e., a sigmoid) is observed in the AIA~94~\AA~images which consists of a set of highly sheared coronal loops (Figure \ref{fig1}(c)). From H$ \alpha $ filtergrams of the active region, we find that a long filament channel exists under the coronal sigmoid which is indicated by arrows in Figure \ref{fig1}(d). We observed significant variations of magnetic flux in AR 11719 from several hours prior to the eruption till the post eruption phase ($ \sim $00:00--10:00 UT). During the eruption of the flux rope, a large M6.5 two-ribbon flare was observed at the location N09E12 together with an associated halo CME. According to the GOES 1--8~\AA~flux, the flare started at 06:55~UT, reached its maximum at 07:16~UT and ended at 7:29~UT. 

\section{Analysis and results}
\label{sec_analysis}
\subsection{Pre-eruption activities}

\subsubsection{Formation and evolution of EUV sigmoid}
\label{pre_eruption_sigmoid}

\begin{figure*}
\plotone{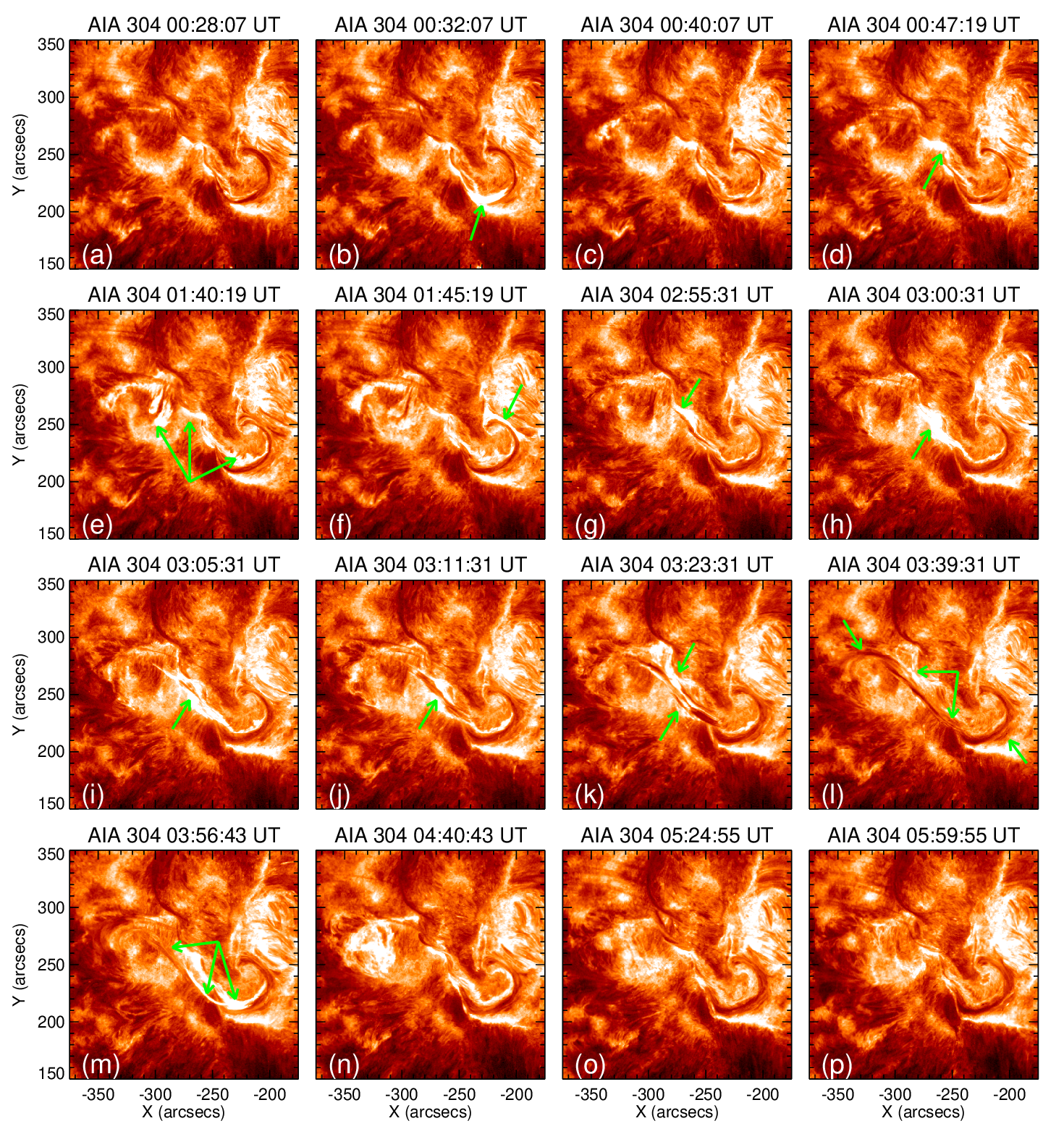}
\caption{Series of AIA 304 \AA~images showing the occurrence of sequential brightenings (indicated by arrows) from different locations of a long filament channel. The emergence of an S-shaped filament can be clearly seen in panel (l) where the relatively straight middle part along with hook-shaped eastern and western portions are indicated by arrows. 
}
\label{fig2}
\end{figure*}

Pre-eruption activities refer to the processes leading to formation and activation of the EUV sigmoid that subsequently erupts during the M6.5 eruptive flare. The evolution of the sigmoid in the pre-eruption phase is illustrated in Figure \ref{Sigmoid_evolution} by a sequence of AIA 94 \AA~images. 
The AIA 94~\AA~channel (Fe~{\footnotesize XVIII}; log(T)=6.8) is apt for the understanding of structures associated with high plasma temperature in the hot flaring corona. In the beginning ($ \sim $00:00 UT), two closely situated bundles of coronal loops are identified (marked by arrows in Figure \ref{Sigmoid_evolution}(a)). At this stage, we cannot clearly identify connectivity between these loops. From $ \sim $1:40 UT, the intensity of the two loop systems increases and we clearly notice the establishment of a connectivity between them in a sequential manner. This phase ($ \sim $1:40--2:00 UT) is characterized by a build-up of bright, diffuse emission in the region that lies between the two loop systems. The coupled loop system undergoes further expansion and the whole region evolves into a large coronal sigmoid at $\sim$ 04:30~UT (see Figure \ref{Sigmoid_evolution}(h)). Further, during the sigmoid formation, the western part of the active region remains active in the form of continued episodic brightening and diffuse emission (shown inside the dotted box in Figure \ref{Sigmoid_evolution}(c)). This region is densely occupied with a cluster of low-lying loops.

It is striking to note the successive emergence of coronal loops in a region that lie between two J-shaped bundles of loops (indicated by arrows in Figure \ref{Sigmoid_evolution}(f)-(g)). Further, the loops in this region brighten up several times ($ \sim $3:00--3:15 UT, $ \sim $3:55--4:10 UT, $ \sim $4:30--4:40 UT) until the full development of the sigmoid structure. 
It is noteworthy that the two J-shaped bundle of loops successive transform into a coherent sigmoid structure via transient loop brightenings that occur between them.

Finally, we highlight the rise of a large bundle of flux rope that evolved into a kinked structure toward the south-east side of the sigmoid (see the region inside the dotted box in Figure~\ref{Sigmoid_evolution}(k)). 
We clearly notice that the top portion of this flux rope exhibits writhing motions with the simultaneous expansion of its legs during which the core of the sigmoid brightens up at multiple locations ($\sim$6:30 UT). We further mention the rise of another thread of this flux rope from the eastern leg of the sigmoid. These rising structures clearly reveal the slow yet steady expansion of a hot flux rope (FR) well before the onset of the impulsive flare emission (See Figure \ref{Sigmoid_evolution}(l)).                

\subsubsection{Episodic energy release in the vicinity of the filament channel}
\label{sec_Episodic_release}

\begin{figure}
\epsscale{1.2}
\plotone{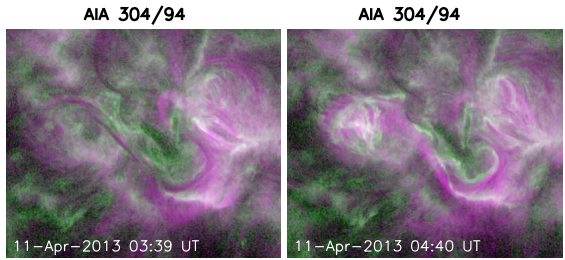}
\caption{Composite of the AIA~94~\AA~(red) and 304~\AA~(green) images. The images reveal an S-shaped filament channel underneath the hot coronal sigmoid. }
\label{aia_composite}
\end{figure}

In Figure \ref{fig2}, we present a sequence of AIA 304 \AA~images of the AR showing the incidences of episodic brightenings in the vicinity of a long filament channel during the pre-eruption phase (i.e., between 00:30 UT and 06:00 UT). The AIA 304~\AA~channel (He~{\footnotesize II}; log(T)=4.7) images the solar structures formed at the chromosphere and the transition region. In Figure~\ref{aia_composite}, we provide composites of the AIA 94~\AA~and 304~\AA~images. These images reveal good spatial correlation between the filament channel and overlying hot coronal sigmoid. The AIA 304~\AA~images reveal interesting evolutionary stages of the filament channels. At the very beginning (i.e., at $\sim$00:28 UT), we observed a U-shaped filament at the south-west part of the AR with localized brightening at its southern side (marked by an arrow in Figure \ref{fig2}(b)). Thereafter, we note episodic brightenings from various portions of the filament channel (marked by arrows in Figures \ref{fig2}(d)-(h)) and simultaneous extension in the length of filament toward the north-east part of the AR. A small portion of the filament, situated at its northern side, undergoes confined eruption and is associated with an intense EUV brightenings at $ \sim $03:05 UT (marked by arrow in Figure \ref{fig2}(i)). At around 03:40 UT, the filament channel attains its maximum length and a clear S-shaped structure emerges (see Figure \ref{fig2}(l)) which is the chromospheric counterpart of the coronal sigmoid seen in AIA 94 \AA~images (Section \ref{pre_eruption_sigmoid}). After complete development of the filament, its eastern part started lifting up and partially disrupted in next few minutes. At this stage, we observed ribbon like brightenings below the rising portion of the filament (see Figure \ref{fig2}(m)). However, the south-west portion of the filament has remained quiet. 

\subsection{Evolution of magnetic flux}
\label{sec_flux_evolution}

\begin{figure}
\plotone{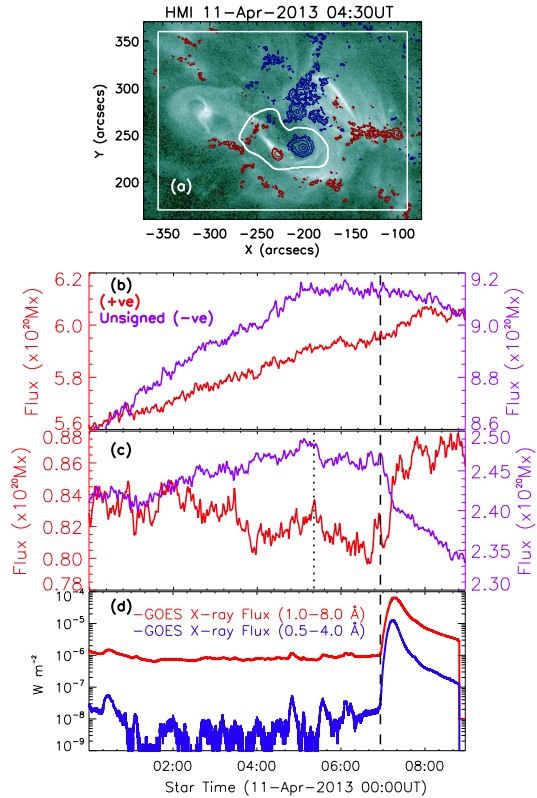}
\caption{Panel (a): HMI LOS magnetogram overplotted on an AIA 94~\AA~image. The positive and negative polarities are shown by red and blue contours, respectively, with contour levels as $\pm$50, $\pm$100, $\pm$150, $\pm$200, $\pm$250~G. This figure also shows the regions of interest for magnetic flux computation: full sigmoidal region (within the rectangular box) and core region (enclosed by the curve at the central part). Panels (b) and (c): The temporal evolution of magnetic flux of positive and negative polarities from the full sigmoidal and core regions, respectively. Panel (d): GOES SXR flux in the 0.5--4 and 1--8 \AA~channels. The vertical dashed line marks the onset of the M6.5 flare. Note a sudden decrease in negative flux in the core flaring region during the rise phase of the flare (panel (c)). The dotted line in panel (c) indicates the time after which both positive and negative flux in the core region undergo a general decrease for $\sim$1.5~hours. Notably, this region is associated with multiple transient loop brightenings during the extended pre-eruption phase as well as the flux rope activation.}
\label{fig4}
\end{figure}

It is well established that the coronal transients are driven by the solar magnetic fields \citep[e.g., reviews by][]{Priest2002,Schrijver2009,Wiegelmann2014} . Therefore, it is crucial to explore how the photospheric magnetic flux evolves prior and during the eruptive phenomena. In Figure~\ref{fig4}(a), we overplot an HMI magnetogram (as contours) displaying the distribution of photospheric line-of-sight magnetic fields on the AIA 94~\AA~image showing the coronal sigmoid. We have selected the following two regions on the HMI magnetogram to investigate the evolution of magnetic flux during the sigmoid-to-arcade transformation: (1) the large region that encompasses the whole sigmoid (see the region defined within the rectangular box in Figure~\ref{fig4}(a)), and (2) the smaller region which forms the central part of the sigmoid (enclosed by a curve in Figure~\ref{fig4}(a)). This central region is of particular interest as transient loop brightenings continuously occurred in this region during which 
the two J-like bundles of loops successively transform into the sigmoid (discussed in Section~\ref{pre_eruption_sigmoid}; also see Figure~\ref{Sigmoid_evolution}(f)-(g)). Notably, during the M6.5 flare, the HXR footpoint emission also occurs within this region, implying this to be the core region associated with magnetic field lines involved in the large-scale magnetic reconnection process (see Section~\ref{sec_M6.5}).            
The positive and negative flux through the extended activity site and the smaller core region are plotted in Figures~\ref{fig4}(b) and (c), respectively. HMI provides the line-of-sight magnetogram at a cadence of 45 s with
a spatial resolution of 0$\arcsec $.5 pixel$ ^{-1}$. We processed the line-of-sight magnetograms for 10~hr starting from 00:00~UT on 2013 April 11. 
The magnetograms are differentially rotated to a common heliographic location and corrected for the line-of-sight effect by multiplying 1/cos($\theta$),
where $\theta$ is the heliocentric angle. We averaged 4 magnetograms
to reduce the noise level. For a comparison between the magnetic flux evolution and coronal energy release, we plot the GOES 1.0--8.0 and 0.5--4.0~\AA~flux in Figure~\ref{fig4}(d). 

We find that the photospheric magnetic fluxes of positive and negative polarities undergo a continuous increase within the sigmoidal active region 
during the pre-eruption phase, i.e., 00:00~UT to 06:55~UT 
(Figure~\ref{fig4}(b)). Notably, this whole duration is characterized by significant coronal activities in the forms of localized episodes of energy release and evolution of bright loops from the EUV sigmoidal region. Further, we note that the positive flux continues to rise following the flare onset at 06:55~UT while the negative flux maintains an almost steady level from $\sim$05:20~UT till the decay of flare emission at $\sim$07:30~UT. More interesting variations of magnetic flux are observed in the core region (Figure~\ref{fig4}(c)). The negative flux undergoes a slow rise until $\sim$05:20~UT while the positive flux exhibits both increasing as well as decreasing episodes. In the later phases ($>$05:20~UT), the negative flux, in general, decreases; Interestingly, the rate of decrease of negative flux is higher during the rise phase of the M6.5 flare. On the other hand, the evolution of positive flux is rather striking. We note that the positive flux decreases during the pre-flare period ($\sim$05:20--06:40~UT). However, it changes its trend completely with the onset of the M6.5 flare and starts to increase.     

\subsection{The eruptive M6.5 flare}
\label{sec_M6.5}
\subsubsection{Flare light curves}
\label{sec_lc}

\begin{figure}
\plotone{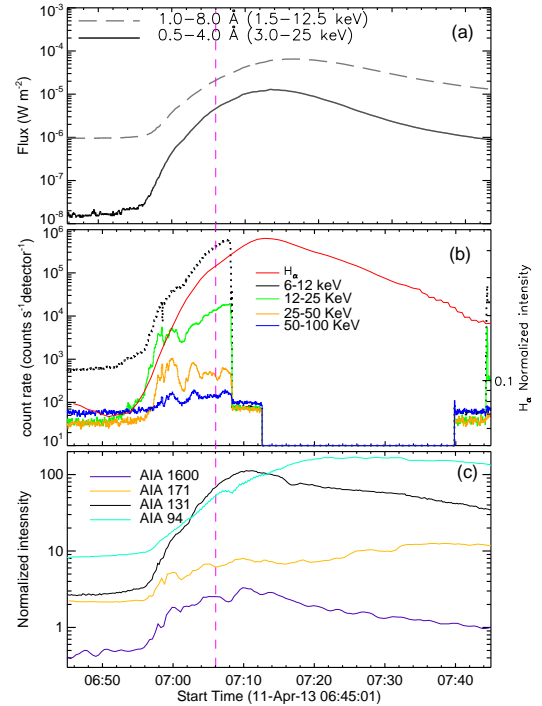}
\caption{X-ray and EUV lightcurves of the M6.5 flare on 2013 April 11. Top panel: GOES SXR profiles at both 0.5--4 and 1--8~\AA~energy channels. Middle panel: RHESSI HXR lightcurves in 4 different energy bands along with GONG H$ \alpha $ intensity profile of the flaring region. Bottom panel: EUV lightcurves at 1600, 171, 131, and 94 \AA~channels. A vertical dashed line (at 07:06 UT) is drawn to differentiate between early rise phase and late rise phase of the flare.}
\label{fig5}
\end{figure}

After a significant pre-eruption activities in the forms of photospheric 	magnetic changes, formation, evolution and activation of a long filament channel, and localized brightenings from various locations close to filament channel, a large M6.5 flare occurred in AR 11719 which is associated with a fast halo CME. In Figure \ref{fig5}, we present the flare light curves at multiple X-ray energy channels from 06:45 to 07:45 UT on 2013 April 11. The GOES SXR profiles presented in Figure \ref{fig5} clearly indicate the onset of flare emission at $\sim $6:55 UT while gradually declining SXR flux lasted until $>$7:40~UT. It is important to note a rather prolonged rise phase of $ \sim $21 minute (from 06:55~UT to 07:16~UT). RHESSI observed this prolonged rise phase but missed subsequent phases (from $ \sim $07:06 to 07:45 UT) due to satellite night-time. In Figure \ref{fig5}(b), we provide RHESSI light curves in different energy bands (6-12, 12-25, 25-50, and 50--100 keV). This figure also presents normalized the H$\alpha$ time profile of the flare showing the average intensity enhancement of the flaring region with respect to the non-flaring background. For this purpose, we have analyzed uninterrupted series of H$\alpha$ filtergrams observed from the Global Oscillation Network Group (GONG). We find that during the prolonged SXR rise phase, multiple high energy HXR peaks ($>$12 keV) are observed (see Figure~\ref{fig5}(b)). In Figure~\ref{fig5}(c), E(UV) light curves of the flare at 94, 131, 171, and 1600~\AA~passbands are shown. 
 
\subsubsection{Sigmoid-to-arcade evolution}

\begin{figure*}
\plotone{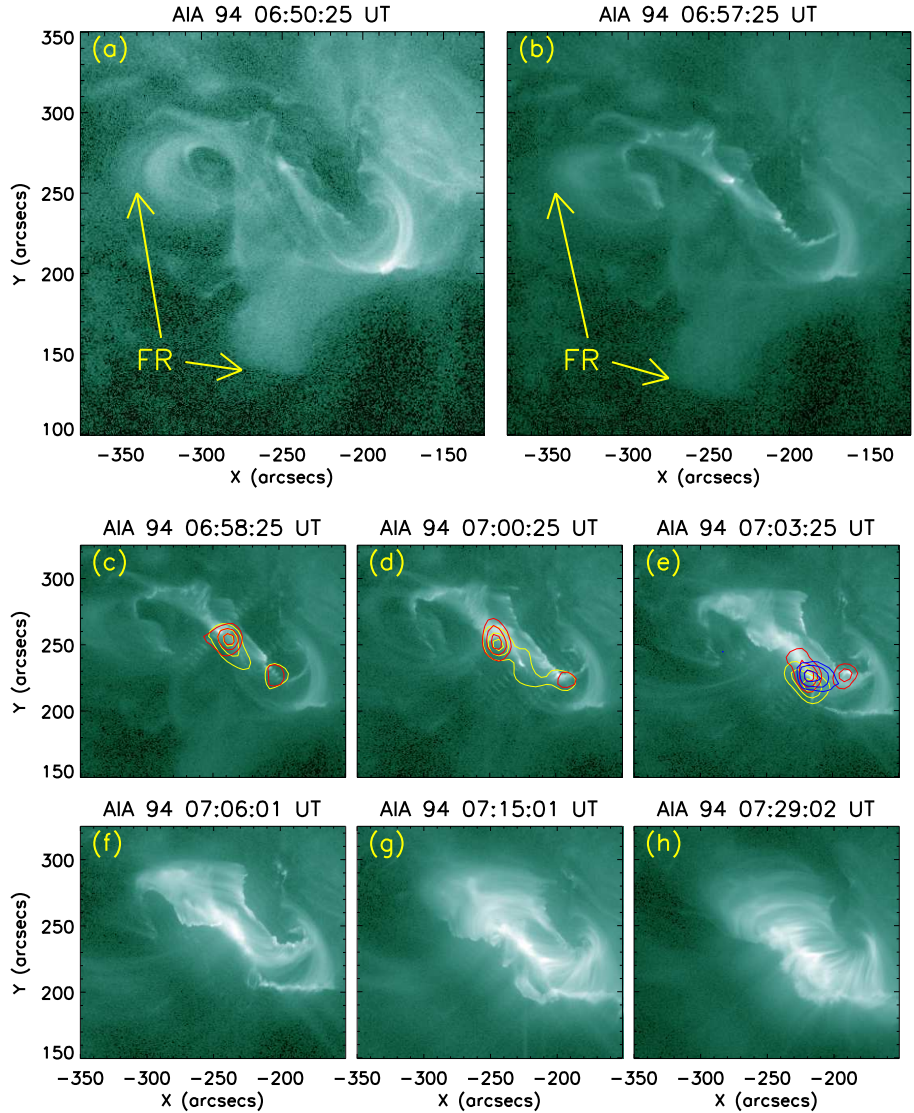}
\vspace{-1.0cm}
\caption{Sequence of AIA 94 \AA~images showing expansion and eruption of the flux rope (FR) during M6.5 flare. Co-temporal RHESSI HXR sources in 12--25 keV (yellow), 25--50 keV (red), and 50--100 keV (blue) are also overplotted on the representative AIA 94 \AA~images (panels (c)--(e)). A bright post-flare loop arcade is formed after eruption of the flux rope (see panels (f)--(h)). RHESSI images are reconstructed with CLEAN algorithm with 1 min integration time. The contour levels are set as 50\%, 70\% and 90\% of the peak flux in each image. (An animation of AIA~94~\AA~observations is available online at ApJ website).}
\label{fig6}
\end{figure*}

\begin{figure*}
\plotone{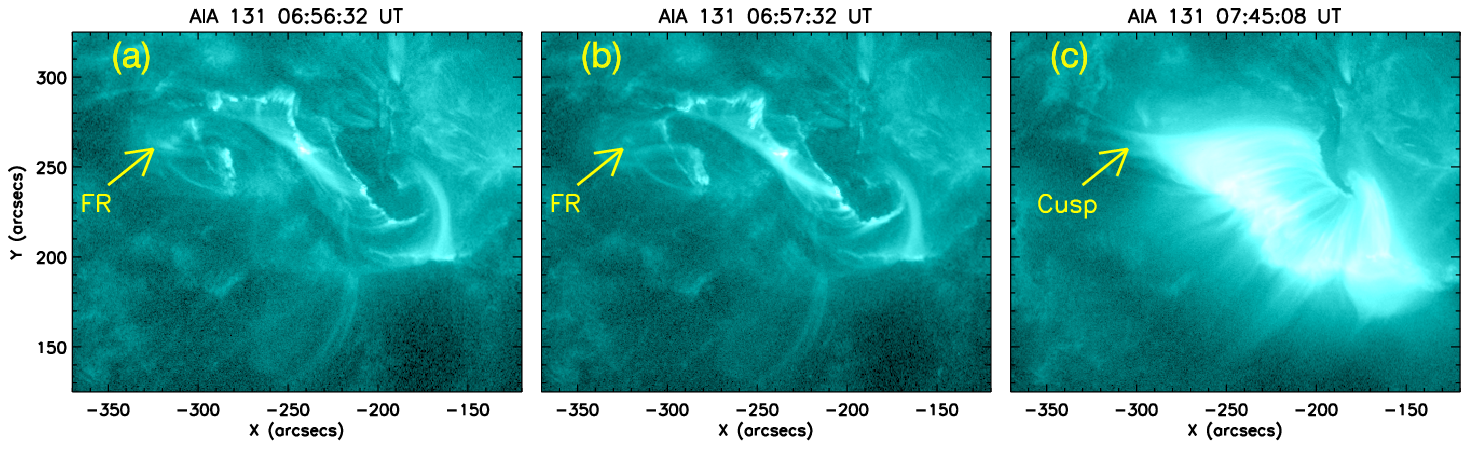}
\caption{Representative AIA 131 \AA~images showing the important stages of the eruption. After the eruption of the flux rope (FR), the flaring region is enveloped by a bright loop arcade. Following the eruption, we note a cusp at the northern end of the loop arcade (panel (c)).}
\label{fig_aia131}
\end{figure*}

In Figure \ref{fig6}, we present series of AIA 94 \AA~images to show the sequential evolution of the sigmoid into the post-flare arcade during various phases of the energy release in the M6.5 flare. The RHESSI HXR contours in different energy bands (12--25 keV: yellow, 25--50 keV: red, and 50--100 keV: blue) are also over plotted on selected EUV images. We recall that the flux rope ascends from $\sim$06:00 UT, i.e., about an hour before the onset of the M6.5 flare (Section \ref{pre_eruption_sigmoid}). We note that the rising flux rope remained in a quasi-stationary state for several minutes ($\sim $6:30--6:55 UT) before undergoing rapid expansion from $ \sim $6:55 UT which marks the impulsive rise of the flare emission (Figures \ref{fig5} and \ref{fig6}). With the rapid eruption of the flux rope, we observe intense brightenings in the source region which subsequently evolved into two distinct flare ribbons. The co-temporal AIA 94 \AA~and RHESSI HXR images reveal that the high energy emission is entirely associated with the middle portion of the EUV sigmoid. Further HXR emission originates in the form of kernels that lie over conjugate EUV flare ribbons. Following the peak phase of SXR emission ($ \sim $7:16 UT; Figure \ref{fig5}), a beautiful system of post-flare arcades envelop the flaring region (Figure~\ref{fig6}(h)) that was occupied with the sigmoid in the pre-eruption phase. In Figure~\ref{fig_aia131}, we present a few representative AIA~131~\AA~images of the activity site. 
The AIA 131~\AA~channel (Fe~{\footnotesize VIII, XXI}; log(T)=5.6, 7.0) observes plasma structures in the transition region and the flaring corona. From these images, we clearly notice that following the eruption of the flux rope, a cusp forms at the apex of hot EUV loops (marked by an arrow in Figure~\ref{fig_aia131}(c)). 
 
\subsubsection{Early rise phase and deviation from the standard flare model}
\label{sec_early_rise}

\begin{figure*}
\epsscale{.9}
\plotone{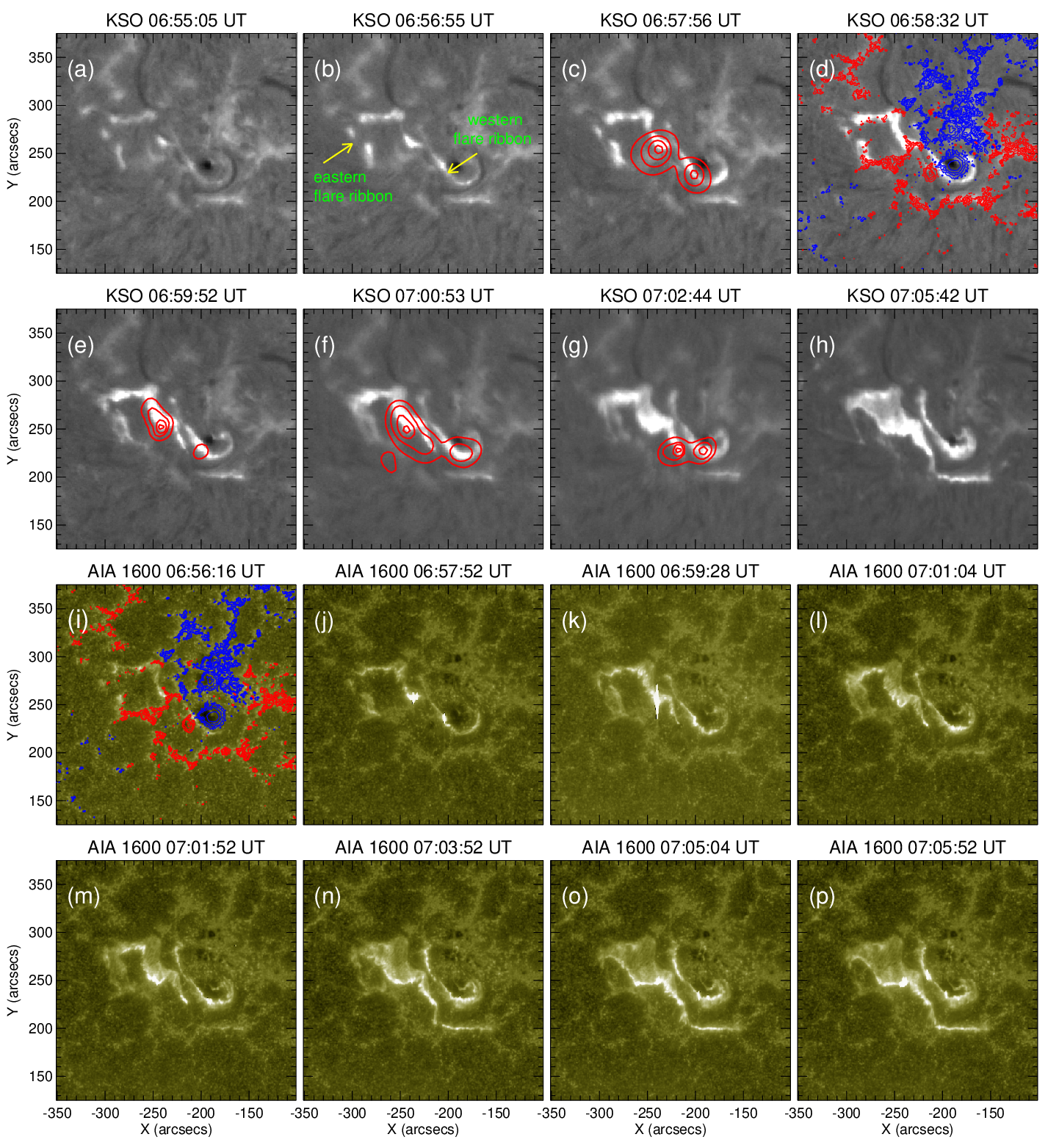}
\caption{KSO H$ \alpha $ (panels (a)-(h)) and AIA 1600 \AA~(panels (i)-(p)) images showing the temporal evolution of flare brightenings/ribbons during early rise phase (i.e., 06:55--07:05 UT; see Figure \ref{fig5}). During this early phase, the flare ribbons present a J-shaped morphology and show lateral extension toward each other. Note that by the end of this early phase ($ \sim $07:05 UT), flare ribbons become parallel to each other and exhibits ``standard" morphology and spatial evolution thereafter. The eastern and western ribbons are indicated by arrows in panel (b). The co-temporal HXR contours in 25--50 keV (red) energy band are overplotted on a few representative H$ \alpha $ images (panels (c), (e)--(g)). In panels (d) and (i), the contours represent the magnetic polarity distribution (blue: negative, red: positive) over H$ \alpha $ and AIA 1600 \AA~images respectively. RHESSI images are reconstructed with PIXON algorithm with 1 min integration time.  The contour levels are set as 10\%, 20\%, 40\%, and 80\% of the peak flux in each image.}
\label{fig7}
\end{figure*}

\begin{figure*}
\plotone{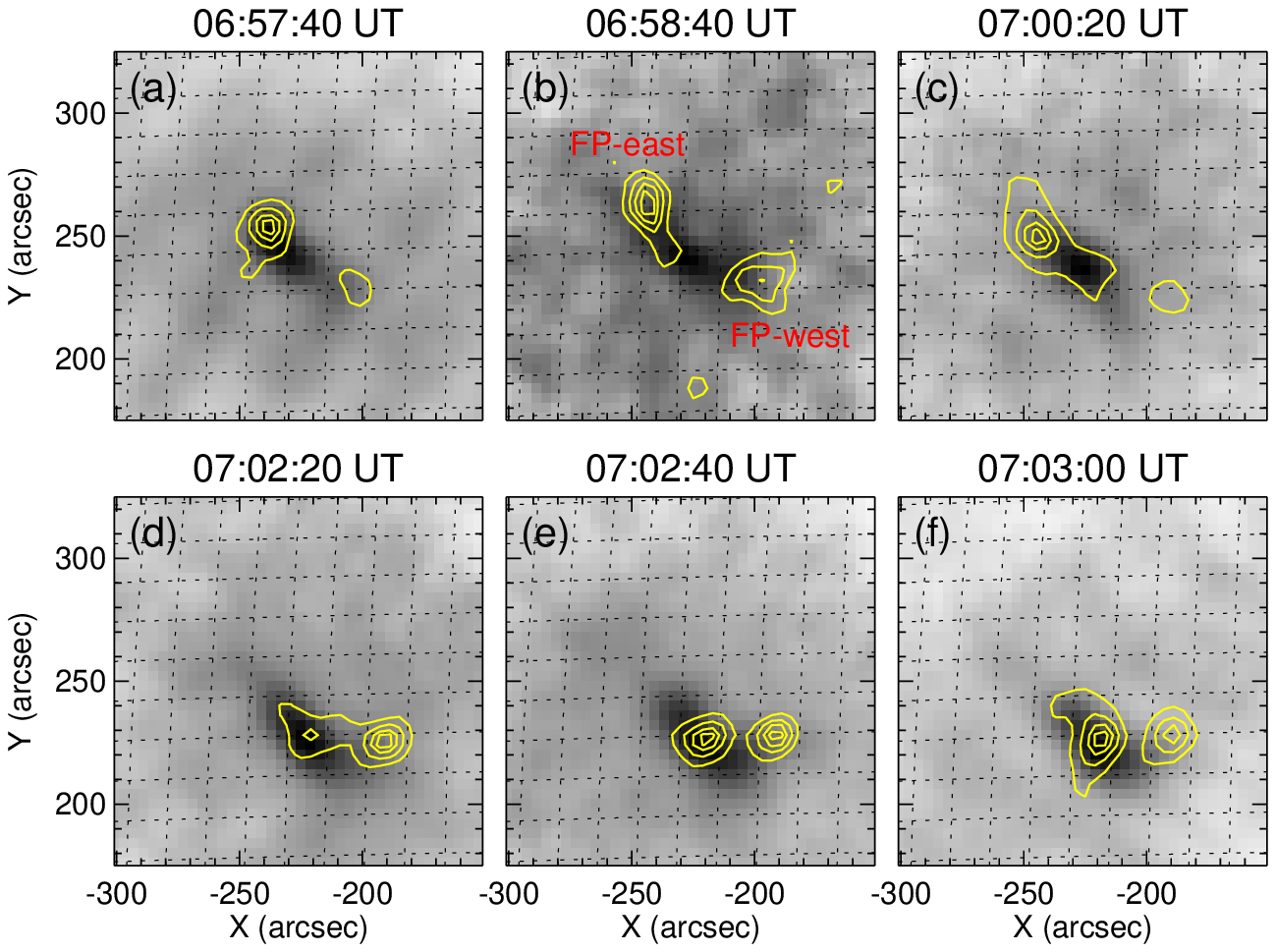}
\caption{Sequence of RHESSI images in 6--12 keV (background grey images) overlaid by co-temporal contours of 25--50 keV HXR sources. We note the converging motion of HXR foot-point sources at 25--50 keV during the prolonged rise phase of the flare. In panel (b), we named the FP sources as per their location as FP-east and FP-west. RHESSI images are reconstructed with CLEAN algorithm with 20 sec integration time.  The contour levels are set as 50\%,           70\%, 85\%, and 95\% of the peak flux in each image}
\label{fig8}
\end{figure*}

From the X-ray and EUV flux profiles, it is evident that the event under study belongs to the category of long duration event (LDE) which are marked by typical evolution of long, parallel flare ribbons in the chromosphere. However, during the prolonged rise phase of this flare with a duration of $ \sim $21 minutes (Figure \ref{fig5}), the flare ribbons and associated HXR sources exhibit a complicated dynamical evolution. For the detailed study of the evolution of flare ribbons, we have presented a sequence of KSO H$ \alpha $ and AIA 1600~\AA~UV images in Figure \ref{fig7}. The sequence of H$\alpha$ filtergrams provide information about the flare morphology in the chromosphere and response of coronal energy release phenomena at this layer. The AIA~1600~{\AA} channel (C~{\footnotesize IV + cont.}; log(T)=5.0) observes combined emission from the transition region and the upper photosphere. The co-temporal HXR contours in 25--50 keV (red) are also overplotted on a few representative H$ \alpha $ images (Figures~\ref{fig7}(c),(e)--(g)). 

The early H$ \alpha $ flare emissions are originated in the form of bright kernels (see Figures \ref{fig7}(a) and (b)) that evolved into the flare ribbons subsequently (marked as the eastern and western flare ribbons in Figure~\ref{fig7}(b)). It is important to note that the flare ribbons exhibit J-shaped structure during the early stages (Figures \ref{fig7}(b)--(f)). In particular, this J-shaped morphology is quite prominent for the eastern flare ribbon which is also larger and undergoes more dynamic evolution during the early rise phase of the flare (i.e., between 06:55 and 07:06 UT). A comparison of H$\alpha$ filtergrams with the evolution of the sigmoid seen in AIA 94 \AA~images (cf. Figures \ref{fig6} and \ref{fig7}) suggests that the hooked part of the J-shape for the eastern flare ribbon is associated and probably physically linked with the eruption of the eastern portion of the overlying flux rope (FR). On the other hand, the western flare ribbon is shorter and exhibits a simple morphology. We note that the HXR conjugate sources in 25--50 keV energy bands nicely correlates with the brightest part of the respective H$\alpha $ flare ribbons (see Figures \ref{fig7}(c),(e)-(g)). In Figures~\ref{fig7}(d) and (i), we have overplotted HMI mangetograms over the co-temporal H$\alpha$ and UV images, respectively, to compare the evolution of flare ribbons with respect to photospheric magnetic polarities. It is noteworthy that the H$\alpha$ filament nicely delineates the separation of opposite magnetic polarities in the photosphere and, therefore, can be considered to approximately outline the  polarity inversion line (PIL). Both H$\alpha$ and UV images clearly reveal that the eastern flare ribbon extends toward south-west direction, parallel to the PIL until $\sim$7:06 UT while the western flare ribbon evolves in the opposite direction. The spatial evolution of flare ribbons observed during the early rise phase is not consistent with the canonical picture of eruptive two-ribbon flares. In particular, the conjugate J-shaped flare ribbons apparently move toward each other parallel to the PIL.

To compare the evolution of low and high energy X-ray emitting regions, we have shown a sequence of RHESSI 6--12 keV images (gray background) overlaid by co-temporal 25--50 keV (yellow contours) in Figure~\ref{fig8}. As described earlier, we observed two distinct HXR sources (marked as FP-east and FP-west in Figure~\ref{fig8}(b)). Due to the limited observations from RHESSI, we could investigate the motion of HXR sources only during the initial $\sim $6 minutes of the rise phase. We find that the separation between conjugate HXR sources decreases in the successive images, further confirming the observed converging motions of flare ribbons (see also Figures~\ref{fig7}(c), (e)-(g)). We also note that initially the FP-west is relatively weak and does not show much spatial variations. On the other hand, FP-east is stronger and moves towards the FP-west. Further, the intensity of FP-west increases with the rise phase and becomes comparable to the FP-east by $\sim$07:02~UT (see Figure \ref{fig8}(e) and (f)). Also by this time, both the FP sources come closest to each other and their corresponding flare ribbons became almost parallel (see Figure \ref{fig7}(n)). The 6--12~keV X-ray source exhibits a single structure and moves slowly toward south-west during the rise phase. 

\subsubsection {Standard phase of the eruptive flare}

\begin{figure*}
\plotone{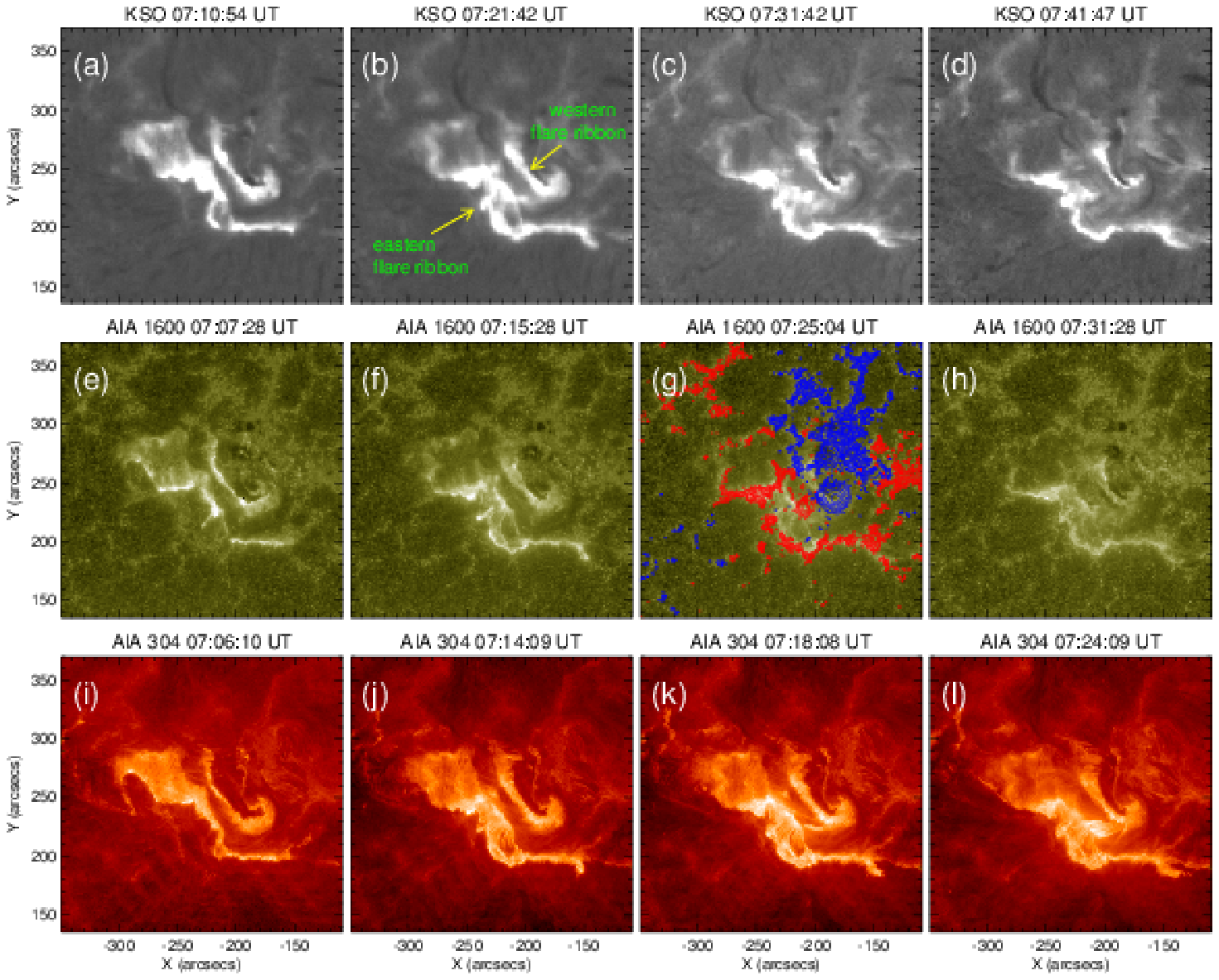}
\caption{The flare evolution during main phase is shown by a sequence of images taken in KSO H$ \alpha $ (top row), AIA 1600 \AA~(middle row) and 304 \AA~(bottom row). The eastern and western ribbons are indicated by arrows in panel (b). The co-temporal magnetogram is also overplotted on a AIA 1600 \AA~image in panel (g). The blue and red contours represent the negative polarity and positive polarity magnetic flux, respectively.}
\label{fig9}
\end{figure*}

The early rise phase (06:55-07:06 UT; see previous section) is characterized by the converging motions of flare ribbons. In the subsequent stages, we note `standard' evolution of a pair of well developed, classical flare ribbons during which they apparently move away from each other in the perpendicular direction to the PIL. To show the standard phase of the M6.5 flare, we present KSO H$\alpha$, AIA~1600~\AA, and AIA~304~\AA~images in Figure~\ref{fig9}. We note that, out of the two ribbons, the eastern ribbon undergoes more dynamic evolution and lateral expansion compared to the western flare ribbon (Figure \ref{fig9}). Notably, the eastern flare ribbon remains prolonged and bright until the late gradual phase of the event while the western flare ribbon appreciably decays in length as well as intensity of the emission. In this context, we emphasize that the western ribbon is associated with stronger magnetic field (negative polarity) while the eastern ribbon forms over the weaker and dispersed flux regions (positive polarity) in the photosphere (Figure \ref{fig9}(g)). These observations indicate crucial evidence for the asymmetric distribution in the injection of accelerated particles into the flare loop systems.
     
\subsection{CME observations}

\begin{figure*}
\epsscale{0.7}
\plotone{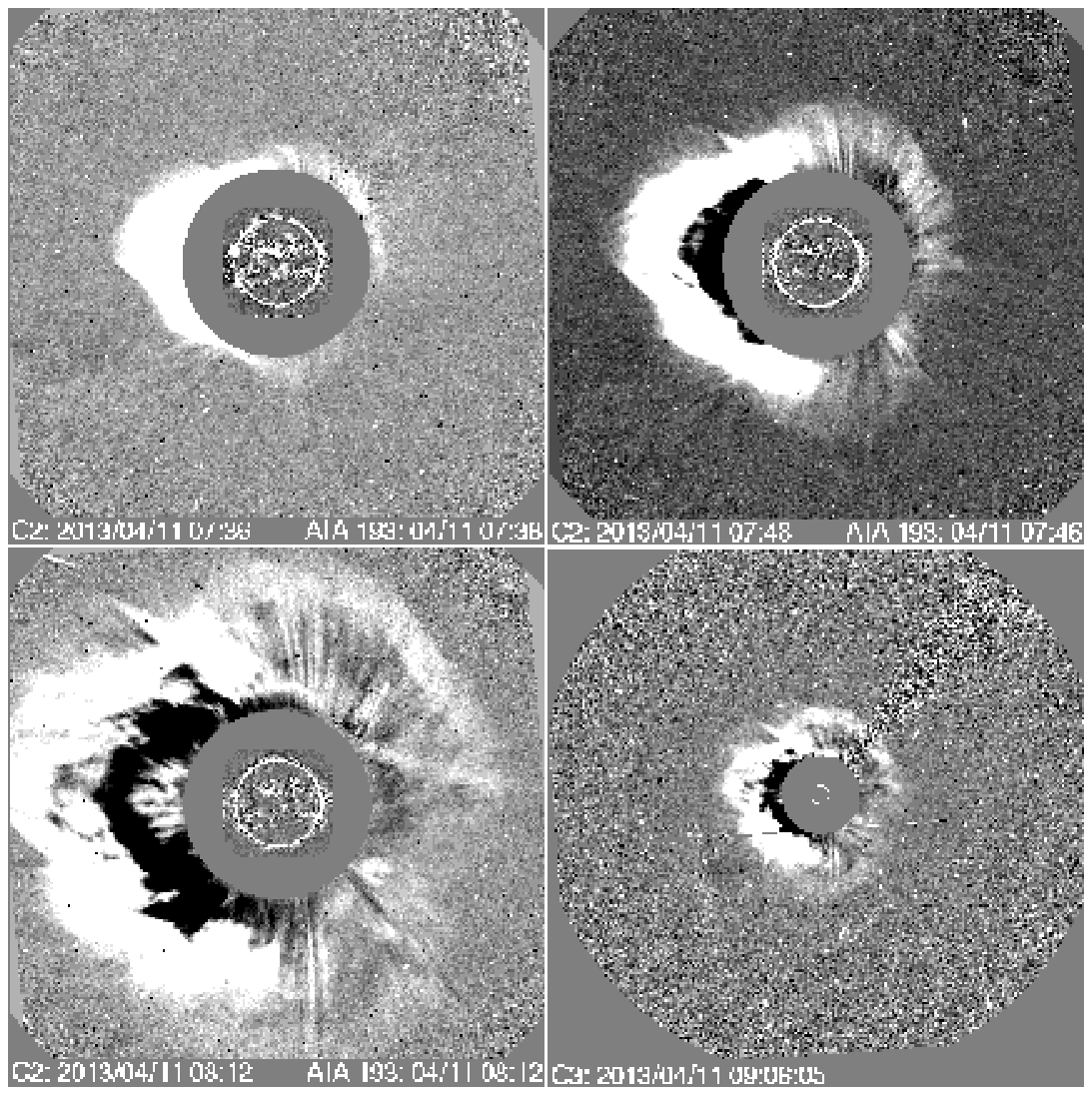}
\caption{Running difference images derived from LASCO C2 and C3 showing the propagation of the CME associated with the M6.5 flare.}
\label{fig10}
\end{figure*} 

The eruption of the flux rope eventually leads to an Earth-directed (halo) CME. In Figure \ref{fig10}, we show a few representative white-light images observed by LASCO on board SOHO. The CME was first seen in C2  field-of-view (FOV) at 07:24 UT at the position angle of 85$^\circ$. The halo structure of the CME emerged after 07:46 UT in the C2 coronagraph which was tracked by C3 coronagraph till 12:00 UT up to a height of 24 R$ _{\odot} $. The height-time plot available at the SOHO LASCO CME catalog\footnote{http:$//cdaw.gsfc.nasa.gov/CME_list/UNIVERSAL/2013_04/univ2013_04.html$} shows that the linear speed of CME is $ \sim $860 km s$ ^{-1} $. Here it should be noted that the CME almost originated in the disk center, thus the linear speed of the CME should be considered as its expansion speed.  A second-order fit to the height-time data indicates a deceleration of $ \sim $8 m s$^{-2}$ in the propagation of CME. Here it is worth to mention that the COR1 coronagraph on board SECCHI/STEREO-B, which observe the inner corona \citep[1.5--4~R$_{\odot}$;][]{Howard2008} with better temporal resolution (5 min), detected this CME at $ \sim $07:10 UT when the M6.5 flare was still going through its rise phase (Figure \ref{fig5}). The early detection of the CME is due to different viewing angle of STEREO-B with the separation angle of 142$^{\circ}$ from Earth (in heliocentric coordinates) on the day of observation.

\subsection{Dynamic radio spectrum}

The HiRAS spectrograph\footnote{http://sunbase.nict.go.jp/solar/denpa/index.html} operated by the NICT, Japan, observed significant activities at a wide range of frequencies between 30 and 2000 MHz in association with this sigmoid eruption. We present HiRAS spectrum in Figure~\ref{fig11}. At higher frequency range of the spectrum ($\sim$1800--600~MHz), we observe type IV continuum emission  from the early rise ($\sim$06:55 UT) to the peak phase of the flare ($\sim$07:16~UT). It is remarkable to note three prominent patches of type IV continua in the beginning ($\sim$06:55--07:08~UT; marked by arrows in Figure~\ref{fig11}). Notably, this interval correlates with the early rise phase of the flare (see Section~\ref{sec_early_rise}). At this time, the eruption initiates from the sigmoid with the expansion of the flux rope (see Figure~\ref{fig6}(a)-(b)). The first patch is the most intense ($\sim$ 06:55--07:01) and forms within the frequency range of $\sim$1800--600 MHz, evidencing this to be situated at very low coronal heights. We find a gradual decrease in the intensity as well as drift of frequencies toward the lower side for the successive type IV continua which imply the continuous bulging of coronal loop systems associated with the sigmoid during the rise phase. It is noteworthy that during the period of early type IV emission, we also observe a bunch of type III radio bursts ($\sim$06:55--07:01~UT) at much lower frequency region of the spectrum (from $\sim$500 to 50 MHz). 

The rise phase of the M6.5 flare is further characterized by an intense and broad type II radio burst during $\sim$07:01--07:14~UT (see region between the vertical dashed lines shown in Figure~\ref{fig11}). However, the weak emission from this burst can be seen up to 07:20~UT. It may be noted that the type II emission start at frequency of $\sim$165~MHz at $\sim$07:02~UT and extends down to $\sim$35~MHz at $\sim$07:14~UT. However, the spectrogram clearly reveals the fundamental emission of this burst running parallel to the above intense burst at a lower frequency range. Using the Newkirk coronal density model {\citep{Newkirk1961}, we estimate the heliocentric heights of the type II burst to be $\sim$1.5--2.6 R$ _{\odot}$ which corresponds to a speed of $\sim$1090~km~s$^{-1}$. The comparisons between the sigmoid evolution observed in the AIA~94~\AA~images (Section \ref{pre_eruption_sigmoid}) with the radio dynamic spectrum provides more insights into the evolutionary stages associated with the flux rope expansion. It is likely that the rapid expansion of the EUV flux rope would induce magnetic reconnection in the stretched overlying magnetic field lines. The escaping electron beams from the reconnection region would thus provide radio signatures in the form of type III radio bursts. The subsequent type II radio emission implies propagation of the shock wave resulting from the eruptive expansion of the associated fluxrope-CME system. 

\section{Discussions}   
\label{sec_discuss}    

\begin{figure*}
\plotone{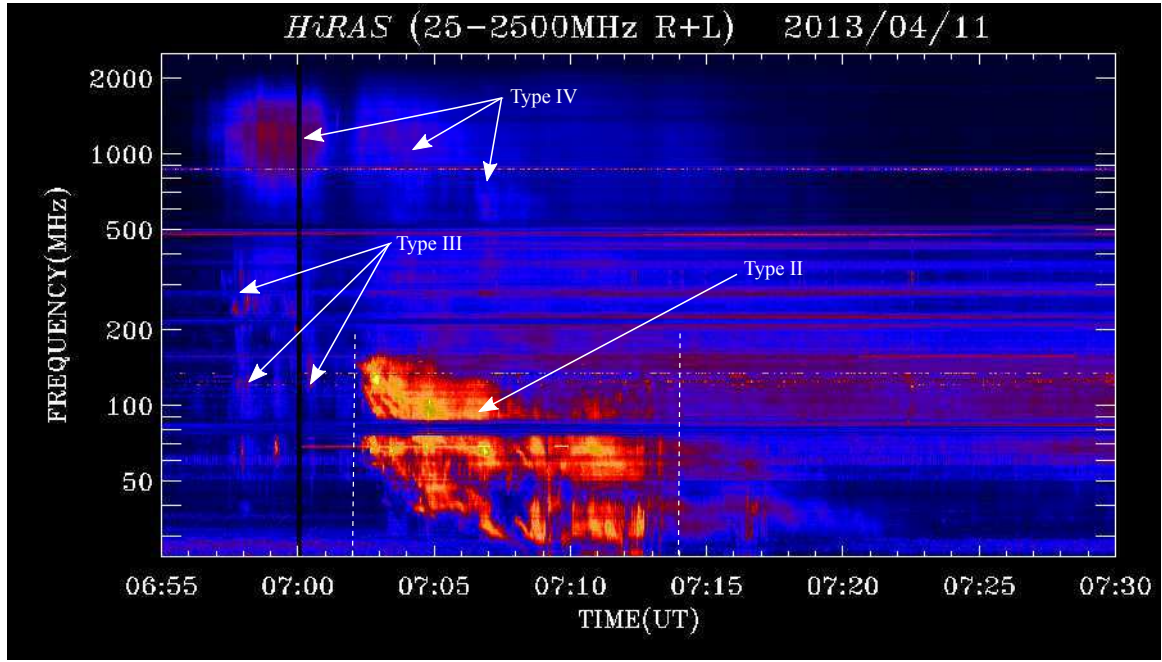}
\caption{Dynamic radio spectrum recorded by the HiRAS spectrograph from 25 to 2500 MHz showing a wide range of radio activities associated with the M6.5 flare/CME event on 11 April 2013. Two vertical dashed lines indicate the start and end timings of type II radio burst considered for the calculation of its speed.}
\label{fig11}
\end{figure*}
      
In this paper, we present a multi-wavelength investigation of a sigmoid-to-arcade development in AR 11719 on 2013 April 11. The study aims to explore several crucial aspects involved during the process of a solar eruption right from the pre-eruption stages of a coronal EUV sigmoid to the post-flare phase. We further provide valuable insights about the physical processes occurring simultaneously at different layers of the solar atmosphere during the successive transformation of hot active region loops (observed in the AIA 94~\AA~images) into the sigmoid that continued over several hours ($\sim$7 hrs) prior to the eruption. 

\subsection{Formation and activation of the sigmoid}

The comprehensive analysis of AIA EUV/UV images and HMI magnetograms of about 9 hours duration was undertaken to explore the formation and eruption of the sigmoid and its association with the basic process of the magnetic flux evolution through the photosphere. The simultaneous observations of the active region in hot (94~\AA; T~$\sim$6~MK) and cool (304~\AA; T~$\sim$50,000~K) channels clearly reveal that the formation of the sigmoid occurred in multiple steps. The EUV observations at 94~\AA~reveal two pre-existing J-shaped coronal loop systems which successively transformed into a sigmoid structure. Coronal sigmoids were discovered in soft X-ray emission as a precursor to CMEs \citep{Rust1996, Manoharan1996}. However, many recent studies have confirmed that sigmoid structures can be observed at different EUV channels which provide evidence that sigmoids exist over a wider range of temperatures \citep{Liu2007,Cheng2014_double_MFR}. Our study reveals that transient and localized brightenings, associated with the successive emergence of hot loops, proceed in the region between the two J-shaped loop bundles as they appear to coalesce and a coherent sigmoid structure evolve (Figure~\ref{Sigmoid_evolution}) evidencing the role of magnetic reconnection toward the formation of the sigmoid. With the complete development of the sigmoid, its extended middle section along with the elbow regions at both sides produce intense emission in the AIA~94~\AA~ images. This implies that the twisted structure is at a higher temperature (T $\sim$6~MK) than the ambient active region corona.     

The phase of sigmoid formation in the corona is associated with dynamical activities at lower layers of the solar atmosphere, viz., chromosphere and transition region. We observe multiple brightenings at different locations of the  filament channel which was observed at 304~\AA~images. A large brightening occurred prior to the merging of two filament channels and the coupled structure continuosly undergo rapid evolution (Figures~\ref{fig2}(i)--(k)). Following multiple, localized events of energy release at different locations, an S-shaped long channel emerged which essentially represents the core region of the overlying EUV sigmoid. The study reveals simultaneous changes in the configuration of overlying as well as core regions during the formation of the sigmoid. The association between the twisted filament structures and overlying coronal sigmoid has been recognized in several studies \citep{Pevtsov1996,Gibson2000,Gibson2002,Pevtsov2002}. In all these papers, the filament was studied in H$\alpha$. \cite{Pevtsov2002} studied a set of active region filaments associated with an X-ray sigmoid. This study reveals that as the eruption proceeds, the sigmoid gets replaced  by a cusp or arcade while the underlying H$\alpha$ filament does not show significant changes. In comparison to the study of \cite{Pevtsov2002}, where filament evolution was studied following the onset of eruption, we have studied the morphological changes in the filament during the formation as well as eruption of the sigmoid. Further, in our study, the filament evolution is studied in EUV images at 304~\AA~that observes structures in transition region and chromosphere which are formed at a higher temperatures than H$\alpha$ features. From these observations, we infer that the chromosphere and its overlying transition region containing dense filament material were highly dynamic when various coronal activities (observed at 94~\AA~channel) occurred while the  sigmoid is taking its shape.
This also suggests that perhaps multiple, localized brightenings in the corona and lower layers are associated with magnetic reconnection at chromospheric and/or transition region heights. \cite{Vemareddy2014} found that the heating and localized brightenings during the activation of the flux rope results in the significant increase of emission measure, density, and temperature of the coronal region, supporting the reconnection scenario. Complementary to these results, \cite{Cheng2015_2MFR} provide evidence from spectroscopic observations from the Interface Region Imaging Spectrograph \citep[IRIS;][]{DePontieu2014} that support the role of magnetic reconnection toward the formation of magnetic flux ropes. In their study, the signatures of magnetic reconnection were observed in the form of red/blue shifts along with non-thermal broadenings at the footpoints of magnetic flux ropes.

Significant variations in photospheric magnetic flux were observed within the sigmoidal region right from its formation stages (Figure~\ref{fig4}(b)).
There was a continuous increase of both positive and negative flux when the whole sigmoidal region is considered (Figure~\ref{fig4}(b)). 
By combining EUV and magnetogram observations, we propose a scenario in which multiple magnetic reconnections, evidenced as localized, transient brightenings, continuously occurs at lower coronal heights with a general increase of photospheric magnetic flux. It is noteworthy that the role of flux emergence has been recognized in simulations toward the build-up of complexity in sigmoids \citep{Archontis2009}. In this model, the continual flux emergence would proceed with the formation of the flux rope through internal reconnections. Thus, the loops which form the overall sigmoid are reconnected field lines heated by reconnection along the current layers.
Our observations indicate that the localized, multiple reconnection events, mainly occurring within the core region, have not only reconfigured the magnetic topology of the region toward the formation of the large sigmoid but also contributed in sustaining the temperature of the sigmoidal flux rope at a higher level than the ambient corona. Although our observations show significant flux emergence during the formation of the flux rope, it may not be common to all the events. For example, \cite{Cheng2014_double_MFR} found no significant flux emergence in the period of magnetic flux rope formation. On the other hand, some studies, pertaining to the long-term evolution of the active regions, have shown significant flux cancellation along the polarity inversion line over a longer time scale ($\sim$2.5$-$3.5 days) during the process of flux rope formation \citep{Green2011,Yardley2016}.

More striking variations of the photospheric magnetic flux were observed from the core field region associated with the flux rope eruption and subsequent M-class flare (see region enclosed by the curve in Figure~\ref{fig4}(a) and corresponding flux plots in Figure~\ref{fig4}(c)). The investigation of flux profiles through this region reveals an increase in negative flux till 05:20~UT. More interestingly, the positive polarity undergoes episodes of increasing as well as decreasing flux evolution. It is worth to recall that during this interval the core region exhibited episodic brightenings characterized by the emergence of a bright loop system. Following each such event, the two J-like bundles of loops successively transform into a more coherent sigmoid structure grows (Figure~\ref{Sigmoid_evolution}).

It is important to note that the flux rope activation has temporal and spatial consistency with the flux cancellation through the core region that started $\sim$1.5 hours before the M6.5 flare (time indicated in Figure~\ref{fig4}(c) by dotted line). We believe that the localized flux cancellation has important implications as the core of the sigmoid encompasses the region of this flux cancellation. We obtain the early signatures of the eruption from the middle and eastern regions of the sigmoid in the form the expansion of bright and twisted structures with diffuse emission, i.e., an ascending flux rope (see coronal structures marked as FR in Figure \ref{fig6}(a)-(b)). It is noteworthy that during this activation phase, we not only observe expansion of the flux rope but also several episodes of localized brightenings from different regions of the sigmoid. We attribute the ascend of the flux rope along with localized brightenings to the slow reconnections driven by the {\it tether-cutting} processes \citep{Moore1992}. It is noteworthy that the main activity center in the large sigmoidal region was its middle portion where the flux cancellation occurred. It is very likely that the {\it tether-cutting} reconnection in the highly sheared core fields, associated with the flux cancellation, initiates the slow ascend of the flux rope. 
Once the flux rope rises to a certain height where the decline of the background magnetic field is fast enough, kink and/or torus instability may set in \citep{Torok2005, Kliem2006} causing its fast eruption and impulsive rise of the M6.5 flare.

\subsection{The M6.5 flare and sigmoid-to-arcade transformation}


The M6.5 flare underwent a very prolonged rise phase of $\sim$21 minutes which is significantly higher than the median rise time of 10 min for M-Class flares \citep{Veronig2002}. The rise phase is characterized by a gradual build up of the SXR flux between 06:55 and 07:16~UT (Figure~\ref{fig5}). A careful analysis of imaging observations at E(UV), H$\alpha$ and HXR channels reveal that the rise phase exhibits a complex dynamical evolution of flare ribbons and HXR footpoints during the early stages. In view of this, we have divided the rise phase of the flare into two parts: early rise phase (06:55--07:06~UT) and late rise phase (07:06--07:16~UT). The fundamental difference of these phases lie in the fact that the emission signatures of the early rise phase do not comply with the scenario of the standard flare model while the late rise phase correlates well with the criteria of the standard flare \citep[see, e.g., ][]{Joshi2012}. Although a flare with prolonged rise phase is subject of interest in itself, it is not an uncommon phenomenon \citep[see, e.g.,][]{Bak-Steslicka2011}.

The morphology and spatial evolution of the flare ribbons during the early rise phase have important implications toward the initiation of the eruption and reconnection. It is noteworthy that the ribbons formed (i.e., brightening in chromosphere started) following fast rise of the flux rope (Figures~\ref{fig6} and \ref{fig7}). In several recent papers, the flux ropes have been identified in images taken at hot EUV channels, viz., 94 and 131~\AA~ \citep{Cheng2011,Patsourakos2013,Kumar2014}. As typically observed, we note flux ropes as a bundle of hot coronal loops in EUV, displaying intense yet diffuse emission. The radio dynamic spectra show an intense patch of type IV in the frequency range of 1800--600~MHz during the activation and eruption of the flux rope (see type IV continua between $\sim$6:55--07:01~UT in Figure~\ref{fig11}). It is likely that this early type IV continuum represents intense emission from the non-thermal electrons trapped in the coronal magnetic structure associated with the sigmoid as the reconnection proceeds with the expansion of the flux rope. In conjunction to the early type IV, we observe a bunch of type III radio bursts during $\sim$06:55--07:01 UT which originates at $\sim$500 MHz and extends up to $\sim$50 MHz. The temporal consistency between the type III and type IV along with combined EUV and HXR images provide a clearer view of the multiple processes occurring simultaneously during the early rise phase. From these multi-wavelength measurements, we infer that with the rise of the flux rope, large-scale magnetic reconnection sets in causing expansion of loop systems and particle acceleration. The type III bursts then imply the ejection of beams of electrons along the open field lines at relativistic speeds from the reconnection region, formed below the erupting flux ropes. The association of type III radio burst with the early stages of the eruption (i.e., activation of flux rope) has been observed in earlier studies \citep[see e.g.,][]{Joshi2007}. Thus, we find that the eruptive expansion of the flux rope provides the earliest signature of the CME in the source region.

The flare ribbons during the early rise phase exhibit a J-shaped morphology (Figure~\ref{fig7}). We particularly emphasize that the eastern ribbon is larger with the hooked part of the ``J"-structure appears to be highly curved. 
It is interesting to note that this region is spatially associated with the flux rope expansion from the middle and eastern portions of the sigmoid (see Figure~\ref{fig6} and \ref{fig7}). Recently, \cite{ChengX2016} investigated the evolution of footprints of erupted magnetic flux ropes from sigmoidal active regions in four events. Their study reveals a common pattern in which the early chromoshepric brightenings are located at the two footpoints, as well as in the regions below the two elbows of the magnetic flux ropes which subsequently evolved into double J-shaped ribbons with the two hooks at the opposite ends 
corresponding to the extended footprints. Further, \cite{Zhao2016} studied the magnetic topology of a sigmoid active region before a major eruption and compared it with the morphology of the subsequently developed flare ribbons. Their study reveals that the morphology of flare ribbons showing hooked or J-shaped structures during the early stages matches with the footprints of the quasi-separatrix layer \citep[QSL;][]{Pariat2012,Demoulin1996}. Notably, the association of J-shaped ribbons with the footprints of QSLs is consistent with the extended standard model of solar flares in 3D \citep{Aulanier2012,Janvier2013}. Thus, we believe that the shape and location of the flare riboons in the chromosphere during the early phases of the eruption have important implications on the geometry of the erupting flux rope and the coronal reconnection sites.

Another important aspect of the flare ribbons during the early rise phase is their converging motions which essentially mean that they 
apparently move parallel to the PIL in the opposite directions (Figure \ref{fig7}). The converging motions of flare ribbons are further supported by the HXR observations that show decrease in the distance between conjugate footpoints (Figure \ref{fig8}). Converging motion of flare ribbons is confirmed by some of the recent observations \citep{Ji2006,Ji2007,Liu2008,Joshi2009} and, thus, it has been identified as an important phenomenon occurring during the rise phase of flares. In particular, \cite{Liu2008} noted that during SOL2002-04-30, the two conjugate HXR footpoints first move toward and then away from each other, mainly parallel and perpendicular to the magnetic inversion line, respectively. Further the transition of these two phases of footpoint motions coincides with the direction reversal of the motion of the looptop source. Explanation of these kind of ribbon motions is still beyond the scope of the standard flare model. 

After a prolonged rise phase, the flare exhibited the features of a ``standard" large eruptive flare as described by the standard flare model. Here the main observational characteristic of the standard flare is recognized in the form of increase in the separation of parallel flare ribbons while they move perpendicular to the PIL (Figure~\ref{fig9}). Following the eruption of flux rope, we observed the development of an arcade of coronal loops over the source region previously occupied by the coronal sigmoid. This phenomenon, called sigmoid-to-arcade transformation, is well established and theoretically studied in relation to the eruptive dynamics of CMEs \citep{Gibson2002, Gibson2004}. 
In the radio spectrum, the violent eruption of the fluxropes is followed by the occurrence of a broad, intense type II radio burst implying propagation of a shock wave associated with the passage of the CME \citep[see e.g.,][]{Cho2005}. We, therefore, conclude that the standard flare model is well applicable to the late phase of this flare. We believe that, during this phase, the main driver of the energy release process is the large-scale magnetic reconnection driven by the erupting flux rope. 

\section{Conclusions} 
\label{sec_conclusion}

Sigmoid-to-arcade development is considered as an important aspect of solar eruptive phenomena. Although the association of sigmoids with eruptions is well recognized, our knowledge is still limited about the formation of these structures and their subsequent activation. In this paper, we provide a comprehensive multi-wavelength investigation of EUV, UV, H$\alpha$, X-ray, radio, and magnetic measurements to probe the evolution of a transient coronal sigmoid in active region NOAA 11719 which activates and erupts leading to a fast halo CME. In the following, we summarize the important results of this study:

\begin{enumerate}
\item
The observations of the active region in the hot EUV channel of AIA (94~\AA) reveal the formation of an S-shaped large flux rope through a sequence of multiple, localized brightenings involving two large pre-existing coronal loop systems. Considering the fact that the bright emission at 94~\AA~images corresponds to hot plasma (T $\sim$6~MK), we attribute the multiple localized brightenings during the loop interactions as repetitive magnetic reconnections. The repetitive reconnection events are temporally and spatially associated with significant variations of the photospheric magnetic flux during the extended pre-eruption period of $\sim$7~hours within the core region associated with the formation of sigmoid. 

\item
During the formation of the large sigmoid structure in the corona, we observe fast morphological evolution of a filament channel in the underlying layers of the chromosphere and the transition region observed in AIA 304~{\AA} images. This region also exhibits localized, transient brightenings in conjunction to the episodic energy release in overlying coronal structures. Thus, we propose that the repetitive magnetic reconnections not only play a key role in the formation of the large sigmoidal flux rope in the corona but also contribute toward sustaining its higher temperature than the ambient coronal structures.

\item
The early ascend of the flux rope is associated with the flux cancellation during $\sim$1.5 hours before the eruptive M6.5 flare. Importantly, the central, core region of the sigmoid encompasses the region of flux cancellation. The eruption initiated in the form of expansion of a flux rope from the core and adjoining {\it elbow} regions of the sigmoid. From these observations, we infer that the localized reconnections, likely driven by the {\it tether-cutting} mechanism, occurring in the highly sheared core fields initiated the early ascend of the flux rope.

\item
The sigmoid eruption leads to a large M6.5 two-ribbon flare. The flare is characterized by a prolonged rise phase of $\sim$21 min in SXR. The flare observation of the early rise phase prominently exhibits deviations from the {\it standard flare model} in terms of converging motions of E(UV) and H$\alpha$ flare ribbons along with the decrease in the separation of HXR footpoints. More importantly, the flare ribbons during this phase presents J-shaped morphology. We find that the hooked part of the J-shaped eastern flare ribbon is spatially and temporally correlated with the the eruption of the overlying flux rope, evidencing a close association between the early morphology of the flare ribbon with configuration of the erupting flux rope.

\end{enumerate} 
 
\acknowledgments
We thank the SDO and RHESSI teams for their open data policy. SDO and RHESSI are NASA's missions under Living With a Star (LWS) and SMall EXplorer (SMEX) programs, respectively. We also acknowledge HiRAS (Hiraiso Radio Spectrograph) operated by the National Institute of Information and Communications Technology (NICT), Japan for providing the dynamic radio spectrum. A.M.V. gratefully acknowledges support from the Austrian Science Fund (FWF): P27292-N20. This work was supported by the BK21 plus program through the National Research Foundation (NRF) funded by the Ministry of Education of Korea. We sincerely thank the anonymous referee for providing constructive comments and suggestions that have significantly enhanced the presentation and quality of the paper.

\bibliographystyle{apj}
\bibliography{ms_references}

\begin{thebibliography}{}
\expandafter\ifx\csname natexlab\endcsname\relax\def\natexlab#1{#1}\fi

\bibitem[{{Archontis} {et~al.}(2009){Archontis}, {Hood}, {Savcheva}, {Golub},
  \& {Deluca}}]{Archontis2009}
{Archontis}, V., {Hood}, A.~W., {Savcheva}, A., {Golub}, L., \& {Deluca}, E.
  2009, \apj, 691, 1276

\bibitem[{{Aulanier} {et~al.}(2012){Aulanier}, {Janvier}, \&
  {Schmieder}}]{Aulanier2012}
{Aulanier}, G., {Janvier}, M., \& {Schmieder}, B. 2012, \aap, 543, A110

\bibitem[{{B{\c a}k-St{\c e}{\'s}licka} {et~al.}(2011){B{\c a}k-St{\c
  e}{\'s}licka}, {Mrozek}, \& {Ko{\l}oma{\'n}ski}}]{Bak-Steslicka2011}
{B{\c a}k-St{\c e}{\'s}licka}, U., {Mrozek}, T., \& {Ko{\l}oma{\'n}ski}, S.
  2011, \solphys, 271, 75

\bibitem[{{Canfield} {et~al.}(1999){Canfield}, {Hudson}, \&
  {McKenzie}}]{Canfield1999}
{Canfield}, R.~C., {Hudson}, H.~S., \& {McKenzie}, D.~E. 1999, \grl, 26, 627

\bibitem[{{Chatterjee} \& {Fan}(2013)}]{Chatterjee2013}
{Chatterjee}, P., \& {Fan}, Y. 2013, \apjl, 778, L8

\bibitem[{{Chen} {et~al.}(2014){Chen}, {Bastian}, \& {Gary}}]{ChenB2014}
{Chen}, B., {Bastian}, T.~S., \& {Gary}, D.~E. 2014, \apj, 794, 149

\bibitem[{{Cheng} \& {Ding}(2016)}]{ChengX2016}
{Cheng}, X., \& {Ding}, M.~D. 2016, \apjs, 225, 16

\bibitem[{{Cheng} {et~al.}(2015){Cheng}, {Ding}, \& {Fang}}]{Cheng2015_2MFR}
{Cheng}, X., {Ding}, M.~D., \& {Fang}, C. 2015, \apj, 804, 82

\bibitem[{{Cheng} {et~al.}(2014{\natexlab{a}}){Cheng}, {Ding}, {Zhang}, {Sun},
  {Guo}, {Wang}, {Kliem}, \& {Deng}}]{Cheng2014_double_MFR}
{Cheng}, X., {Ding}, M.~D., {Zhang}, J., {et~al.} 2014{\natexlab{a}}, \apj,
  789, 93

\bibitem[{{Cheng} {et~al.}(2013{\natexlab{a}}){Cheng}, {Zhang}, {Ding}, {Liu},
  \& {Poomvises}}]{Cheng2013_MRF_CME}
{Cheng}, X., {Zhang}, J., {Ding}, M.~D., {Liu}, Y., \& {Poomvises}, W.
  2013{\natexlab{a}}, \apj, 763, 43

\bibitem[{{Cheng} {et~al.}(2013{\natexlab{b}}){Cheng}, {Zhang}, {Ding},
  {Olmedo}, {Sun}, {Guo}, \& {Liu}}]{ChengX2013_two_MFR}
{Cheng}, X., {Zhang}, J., {Ding}, M.~D., {et~al.} 2013{\natexlab{b}}, \apjl,
  769, L25

\bibitem[{{Cheng} {et~al.}(2011){Cheng}, {Zhang}, {Liu}, \& {Ding}}]{Cheng2011}
{Cheng}, X., {Zhang}, J., {Liu}, Y., \& {Ding}, M.~D. 2011, \apjl, 732, L25

\bibitem[{{Cheng} {et~al.}(2014{\natexlab{b}}){Cheng}, {Ding}, {Guo}, {Zhang},
  {Vourlidas}, {Liu}, {Olmedo}, {Sun}, \& {Li}}]{Cheng2014_MFR_evol}
{Cheng}, X., {Ding}, M.~D., {Guo}, Y., {et~al.} 2014{\natexlab{b}}, \apj, 780,
  28

\bibitem[{{Cho} {et~al.}(2005){Cho}, {Moon}, {Dryer}, {Shanmugaraju}, {Fry},
  {Kim}, {Bong}, \& {Park}}]{Cho2005}
{Cho}, K.-S., {Moon}, Y.-J., {Dryer}, M., {et~al.} 2005, Journal of Geophysical
  Research (Space Physics), 110, A12101

\bibitem[{{De Pontieu} {et~al.}(2014){De Pontieu}, {Title}, {Lemen}, {Kushner},
  {Akin}, {Allard}, {Berger}, {Boerner}, {Cheung}, {Chou}, {Drake}, {Duncan},
  {Freeland}, {Heyman}, {Hoffman}, {Hurlburt}, {Lindgren}, {Mathur}, {Rehse},
  {Sabolish}, {Seguin}, {Schrijver}, {Tarbell}, {W{\"u}lser}, {Wolfson},
  {Yanari}, {Mudge}, {Nguyen-Phuc}, {Timmons}, {van Bezooijen}, {Weingrod},
  {Brookner}, {Butcher}, {Dougherty}, {Eder}, {Knagenhjelm}, {Larsen},
  {Mansir}, {Phan}, {Boyle}, {Cheimets}, {DeLuca}, {Golub}, {Gates}, {Hertz},
  {McKillop}, {Park}, {Perry}, {Podgorski}, {Reeves}, {Saar}, {Testa}, {Tian},
  {Weber}, {Dunn}, {Eccles}, {Jaeggli}, {Kankelborg}, {Mashburn}, {Pust},
  {Springer}, {Carvalho}, {Kleint}, {Marmie}, {Mazmanian}, {Pereira}, {Sawyer},
  {Strong}, {Worden}, {Carlsson}, {Hansteen}, {Leenaarts}, {Wiesmann},
  {Aloise}, {Chu}, {Bush}, {Scherrer}, {Brekke}, {Martinez-Sykora}, {Lites},
  {McIntosh}, {Uitenbroek}, {Okamoto}, {Gummin}, {Auker}, {Jerram}, {Pool}, \&
  {Waltham}}]{DePontieu2014}
{De Pontieu}, B., {Title}, A.~M., {Lemen}, J.~R., {et~al.} 2014, \solphys, 289,
  2733

\bibitem[{{D{\'e}moulin} {et~al.}(1996){D{\'e}moulin}, {Priest}, \&
  {Lonie}}]{Demoulin1996}
{D{\'e}moulin}, P., {Priest}, E.~R., \& {Lonie}, D.~P. 1996, \jgr, 101, 7631

\bibitem[{{Gibson} {et~al.}(2004){Gibson}, {Fan}, {Mandrini}, {Fisher}, \&
  {Demoulin}}]{Gibson2004}
{Gibson}, S.~E., {Fan}, Y., {Mandrini}, C., {Fisher}, G., \& {Demoulin}, P.
  2004, \apj, 617, 600

\bibitem[{{Gibson} {et~al.}(2006){Gibson}, {Fan}, {T{\"o}r{\"o}k}, \&
  {Kliem}}]{Gibson2006}
{Gibson}, S.~E., {Fan}, Y., {T{\"o}r{\"o}k}, T., \& {Kliem}, B. 2006, \ssr,
  124, 131

\bibitem[{{Gibson} \& {Low}(2000)}]{Gibson2000}
{Gibson}, S.~E., \& {Low}, B.~C. 2000, \jgr, 105, 18187

\bibitem[{{Gibson} {et~al.}(2002){Gibson}, {Fletcher}, {Del Zanna}, {Pike},
  {Mason}, {Mandrini}, {D{\'e}moulin}, {Gilbert}, {Burkepile}, {Holzer},
  {Alexander}, {Liu}, {Nitta}, {Qiu}, {Schmieder}, \& {Thompson}}]{Gibson2002}
{Gibson}, S.~E., {Fletcher}, L., {Del Zanna}, G., {et~al.} 2002, \apj, 574,
  1021

\bibitem[{{Glover} {et~al.}(2001){Glover}, {Harra}, {Matthews}, {Hori}, \&
  {Culhane}}]{Glover2001}
{Glover}, A., {Harra}, L.~K., {Matthews}, S.~A., {Hori}, K., \& {Culhane},
  J.~L. 2001, \aap, 378, 239

\bibitem[{{Green} {et~al.}(2011){Green}, {Kliem}, \& {Wallace}}]{Green2011}
{Green}, L.~M., {Kliem}, B., \& {Wallace}, A.~J. 2011, \aap, 526, A2

\bibitem[{{Howard} {et~al.}(2008){Howard}, {Moses}, {Vourlidas}, {Newmark},
  {Socker}, {Plunkett}, {Korendyke}, {Cook}, {Hurley}, {Davila}, {Thompson},
  {St Cyr}, {Mentzell}, {Mehalick}, {Lemen}, {Wuelser}, {Duncan}, {Tarbell},
  {Wolfson}, {Moore}, {Harrison}, {Waltham}, {Lang}, {Davis}, {Eyles},
  {Mapson-Menard}, {Simnett}, {Halain}, {Defise}, {Mazy}, {Rochus}, {Mercier},
  {Ravet}, {Delmotte}, {Auchere}, {Delaboudiniere}, {Bothmer}, {Deutsch},
  {Wang}, {Rich}, {Cooper}, {Stephens}, {Maahs}, {Baugh}, {McMullin}, \&
  {Carter}}]{Howard2008}
{Howard}, R.~A., {Moses}, J.~D., {Vourlidas}, A., {et~al.} 2008, \ssr, 136, 67

\bibitem[{{Hudson} {et~al.}(1998){Hudson}, {Lemen}, {St.~Cyr}, {Sterling}, \&
  {Webb}}]{Hudson1998}
{Hudson}, H.~S., {Lemen}, J.~R., {St.~Cyr}, O.~C., {Sterling}, A.~C., \&
  {Webb}, D.~F. 1998, \grl, 25, 2481

\bibitem[{{Janvier} {et~al.}(2013){Janvier}, {Aulanier}, {Pariat}, \&
  {D{\'e}moulin}}]{Janvier2013}
{Janvier}, M., {Aulanier}, G., {Pariat}, E., \& {D{\'e}moulin}, P. 2013, \aap,
  555, A77

\bibitem[{{Ji} {et~al.}(2007){Ji}, {Huang}, \& {Wang}}]{Ji2007}
{Ji}, H., {Huang}, G., \& {Wang}, H. 2007, \apj, 660, 893

\bibitem[{{Ji} {et~al.}(2006){Ji}, {Huang}, {Wang}, {Zhou}, {Li}, {Zhang}, \&
  {Song}}]{Ji2006}
{Ji}, H., {Huang}, G., {Wang}, H., {et~al.} 2006, \apjl, 636, L173

\bibitem[{{Jiang} \& {Feng}(2016)}]{Jiang2016}
{Jiang}, C.-W., \& {Feng}, X.-S. 2016, Research in Astronomy and Astrophysics,
  16, 015

\bibitem[{{Joshi} {et~al.}(2007){Joshi}, {Manoharan}, {Veronig}, {Pant}, \&
  {Pandey}}]{Joshi2007}
{Joshi}, B., {Manoharan}, P.~K., {Veronig}, A.~M., {Pant}, P., \& {Pandey}, K.
  2007, \solphys, 242, 143

\bibitem[{{Joshi} {et~al.}(2012){Joshi}, {Veronig}, {Manoharan}, \&
  {Somov}}]{Joshi2012}
{Joshi}, B., {Veronig}, A., {Manoharan}, P.~K., \& {Somov}, B.~V. 2012,
  Astrophysics and Space Science Proceedings, 33, 29

\bibitem[{{Joshi} {et~al.}(2009){Joshi}, {Veronig}, {Cho}, {Bong}, {Somov},
  {Moon}, {Lee}, {Manoharan}, \& {Kim}}]{Joshi2009}
{Joshi}, B., {Veronig}, A., {Cho}, K.-S., {et~al.} 2009, \apj, 706, 1438

\bibitem[{{Kliem} {et~al.}(2004){Kliem}, {Titov}, \&
  {T{\"o}r{\"o}k}}]{Kliem2004}
{Kliem}, B., {Titov}, V.~S., \& {T{\"o}r{\"o}k}, T. 2004, \aap, 413, L23

\bibitem[{{Kliem} \& {T{\"o}r{\"o}k}(2006)}]{Kliem2006}
{Kliem}, B., \& {T{\"o}r{\"o}k}, T. 2006, Physical Review Letters, 96, 255002

\bibitem[{{Kumar} \& {Cho}(2014)}]{Kumar2014}
{Kumar}, P., \& {Cho}, K.-S. 2014, \aap, 572, A83

\bibitem[{{Kumar} {et~al.}(2016){Kumar}, {Bhattacharyya}, {Joshi}, \&
  {Smolarkiewicz}}]{Kumar2016}
{Kumar}, S., {Bhattacharyya}, R., {Joshi}, B., \& {Smolarkiewicz}, P.~K. 2016,
  \apj, 830, 80

\bibitem[{{Lemen} {et~al.}(2012){Lemen}, {Title}, {Akin}, {Boerner}, {Chou},
  {Drake}, {Duncan}, {Edwards}, {Friedlaender}, {Heyman}, {Hurlburt}, {Katz},
  {Kushner}, {Levay}, {Lindgren}, {Mathur}, {McFeaters}, {Mitchell}, {Rehse},
  {Schrijver}, {Springer}, {Stern}, {Tarbell}, {Wuelser}, {Wolfson}, {Yanari},
  {Bookbinder}, {Cheimets}, {Caldwell}, {Deluca}, {Gates}, {Golub}, {Park},
  {Podgorski}, {Bush}, {Scherrer}, {Gummin}, {Smith}, {Auker}, {Jerram},
  {Pool}, {Soufli}, {Windt}, {Beardsley}, {Clapp}, {Lang}, \&
  {Waltham}}]{Lemen2012}
{Lemen}, J.~R., {Title}, A.~M., {Akin}, D.~J., {et~al.} 2012, \solphys, 275, 17

\bibitem[{{Lin} {et~al.}(2002){Lin}, {Dennis}, {Hurford}, {Smith}, {Zehnder},
  {Harvey}, {Curtis}, {Pankow}, {Turin}, {Bester}, {Csillaghy}, {Lewis},
  {Madden}, {van Beek}, {Appleby}, {Raudorf}, {McTiernan}, {Ramaty}, {Schmahl},
  {Schwartz}, {Krucker}, {Abiad}, {Quinn}, {Berg}, {Hashii}, {Sterling},
  {Jackson}, {Pratt}, {Campbell}, {Malone}, {Landis}, {Barrington-Leigh},
  {Slassi-Sennou}, {Cork}, {Clark}, {Amato}, {Orwig}, {Boyle}, {Banks},
  {Shirey}, {Tolbert}, {Zarro}, {Snow}, {Thomsen}, {Henneck}, {McHedlishvili},
  {Ming}, {Fivian}, {Jordan}, {Wanner}, {Crubb}, {Preble}, {Matranga}, {Benz},
  {Hudson}, {Canfield}, {Holman}, {Crannell}, {Kosugi}, {Emslie}, {Vilmer},
  {Brown}, {Johns-Krull}, {Aschwanden}, {Metcalf}, \& {Conway}}]{LinRP2002}
{Lin}, R.~P., {Dennis}, B.~R., {Hurford}, G.~J., {et~al.} 2002, \solphys, 210,
  3

\bibitem[{{Liu} {et~al.}(2007){Liu}, {Lee}, {Yurchyshyn}, {Deng}, {Cho},
  {Karlick{\'y}}, \& {Wang}}]{Liu2007}
{Liu}, C., {Lee}, J., {Yurchyshyn}, V., {et~al.} 2007, \apj, 669, 1372

\bibitem[{{Liu} {et~al.}(2008){Liu}, {Petrosian}, {Dennis}, \&
  {Jiang}}]{Liu2008}
{Liu}, W., {Petrosian}, V., {Dennis}, B.~R., \& {Jiang}, Y.~W. 2008, \apj, 676,
  704

\bibitem[{{Manoharan} {et~al.}(1996){Manoharan}, {van Driel-Gesztelyi}, {Pick},
  \& {Demoulin}}]{Manoharan1996}
{Manoharan}, P.~K., {van Driel-Gesztelyi}, L., {Pick}, M., \& {Demoulin}, P.
  1996, \apjl, 468, L73

\bibitem[{{McKenzie} \& {Canfield}(2008)}]{McKenzie2008}
{McKenzie}, D.~E., \& {Canfield}, R.~C. 2008, \aap, 481, L65

\bibitem[{{Moore} \& {Roumeliotis}(1992)}]{Moore1992}
{Moore}, R.~L., \& {Roumeliotis}, G. 1992, in Lecture Notes in Physics, Berlin
  Springer Verlag, Vol. 399, IAU Colloq. 133: Eruptive Solar Flares, ed.
  Z.~{Svestka}, B.~V. {Jackson}, \& M.~E. {Machado}, 69

\bibitem[{{Moore} {et~al.}(2001){Moore}, {Sterling}, {Hudson}, \&
  {Lemen}}]{Moore2001}
{Moore}, R.~L., {Sterling}, A.~C., {Hudson}, H.~S., \& {Lemen}, J.~R. 2001,
  \apj, 552, 833

\bibitem[{{Newkirk}(1961)}]{Newkirk1961}
{Newkirk}, Jr., G. 1961, \apj, 133, 983

\bibitem[{{Pariat} \& {D{\'e}moulin}(2012)}]{Pariat2012}
{Pariat}, E., \& {D{\'e}moulin}, P. 2012, \aap, 541, A78

\bibitem[{{Patsourakos} {et~al.}(2013){Patsourakos}, {Vourlidas}, \&
  {Stenborg}}]{Patsourakos2013}
{Patsourakos}, S., {Vourlidas}, A., \& {Stenborg}, G. 2013, \apj, 764, 125

\bibitem[{{Pevtsov}(2002)}]{Pevtsov2002}
{Pevtsov}, A.~A. 2002, \solphys, 207, 111

\bibitem[{{Pevtsov} {et~al.}(1996){Pevtsov}, {Canfield}, \&
  {Zirin}}]{Pevtsov1996}
{Pevtsov}, A.~A., {Canfield}, R.~C., \& {Zirin}, H. 1996, \apj, 473, 533

\bibitem[{{P{\"o}tzi} {et~al.}(2015){P{\"o}tzi}, {Veronig}, {Riegler},
  {Amerstorfer}, {Pock}, {Temmer}, {Polanec}, \& {Baumgartner}}]{Potzi2015}
{P{\"o}tzi}, W., {Veronig}, A.~M., {Riegler}, G., {et~al.} 2015, \solphys, 290,
  951

\bibitem[{{Priest} \& {Forbes}(2002)}]{Priest2002}
{Priest}, E.~R., \& {Forbes}, T.~G. 2002, \aapr, 10, 313

\bibitem[{{Rust} \& {Kumar}(1996)}]{Rust1996}
{Rust}, D.~M., \& {Kumar}, A. 1996, \apjl, 464, L199

\bibitem[{{Schmieder} {et~al.}(2015){Schmieder}, {Aulanier}, \& {Vr{\v
  s}nak}}]{Schmieder2015}
{Schmieder}, B., {Aulanier}, G., \& {Vr{\v s}nak}, B. 2015, \solphys, 290, 3457

\bibitem[{{Schou} {et~al.}(2012){Schou}, {Scherrer}, {Bush}, {Wachter},
  {Couvidat}, {Rabello-Soares}, {Bogart}, {Hoeksema}, {Liu}, {Duvall}, {Akin},
  {Allard}, {Miles}, {Rairden}, {Shine}, {Tarbell}, {Title}, {Wolfson},
  {Elmore}, {Norton}, \& {Tomczyk}}]{Schou2012}
{Schou}, J., {Scherrer}, P.~H., {Bush}, R.~I., {et~al.} 2012, \solphys, 275,
  229

\bibitem[{{Schrijver}(2009)}]{Schrijver2009}
{Schrijver}, C.~J. 2009, Advances in Space Research, 43, 739

\bibitem[{{Sterling} \& {Hudson}(1997)}]{Sterling1997}
{Sterling}, A.~C., \& {Hudson}, H.~S. 1997, \apjl, 491, L55

\bibitem[{{Sterling} {et~al.}(2000){Sterling}, {Hudson}, {Thompson}, \&
  {Zarro}}]{Sterling2000}
{Sterling}, A.~C., {Hudson}, H.~S., {Thompson}, B.~J., \& {Zarro}, D.~M. 2000,
  \apj, 532, 628

\bibitem[{{Titov} \& {D{\'e}moulin}(1999)}]{Titov1999}
{Titov}, V.~S., \& {D{\'e}moulin}, P. 1999, \aap, 351, 707

\bibitem[{{T{\"o}r{\"o}k} \& {Kliem}(2005)}]{Torok2005}
{T{\"o}r{\"o}k}, T., \& {Kliem}, B. 2005, \apjl, 630, L97

\bibitem[{{Vemareddy} \& {Zhang}(2014)}]{Vemareddy2014}
{Vemareddy}, P., \& {Zhang}, J. 2014, \apj, 797, 80

\bibitem[{{Veronig} {et~al.}(2002){Veronig}, {Temmer}, {Hanslmeier}, {Otruba},
  \& {Messerotti}}]{Veronig2002}
{Veronig}, A., {Temmer}, M., {Hanslmeier}, A., {Otruba}, W., \& {Messerotti},
  M. 2002, \aap, 382, 1070

\bibitem[{{Wiegelmann} {et~al.}(2014){Wiegelmann}, {Thalmann}, \&
  {Solanki}}]{Wiegelmann2014}
{Wiegelmann}, T., {Thalmann}, J.~K., \& {Solanki}, S.~K. 2014, \aapr, 22, 78

\bibitem[{{Yardley} {et~al.}(2016){Yardley}, {Green}, {Williams}, {van
  Driel-Gesztelyi}, {Valori}, \& {Dacie}}]{Yardley2016}
{Yardley}, S.~L., {Green}, L.~M., {Williams}, D.~R., {et~al.} 2016, \apj, 827,
  151

\bibitem[{{Zhao} {et~al.}(2016){Zhao}, {Gilchrist}, {Aulanier}, {Schmieder},
  {Pariat}, \& {Li}}]{Zhao2016}
{Zhao}, J., {Gilchrist}, S.~A., {Aulanier}, G., {et~al.} 2016, \apj, 823, 62

\end{thebibliography}


\end{document}